\documentclass{aa}  

\usepackage{graphicx}
\usepackage{txfonts}
\usepackage{hyperref}
\begin{document}

   \title{Thermal solutions of strongly magnetized disks and the hysteresis in X-ray binaries}

   \author{Nicolas Scepi\inst{1,2,3},  Jason Dexter\inst{3,4}, Mitchell C. Begelman\inst{3,4}, Grégoire Marcel\inst{5,6}, Jonathan Ferreira\inst{1} \and Pierre-Olivier Petrucci\inst{1}}
 \authorrunning{N. Scepi et al.}
   \institute{Univ. Grenoble Alpes, CNRS, IPAG, 38000 Grenoble, France
\and School of Physics and Astronomy, University of Southampton, Highfield, Southampton, SO17 1BJ, UK
\and JILA, University of Colorado and National Institute of Standards and Technology, 440 UCB, Boulder, CO 80309-0440, USA 
\and Department of Astrophysial and Planetary Sciences, University of Colorado, 391 UCB, Boulder, CO 80309-0391, USA
\and Institute of Astronomy, University of Cambridge, Madingley Road, Cambridge CB3 OHA, UK 
\and Nicolaus Copernicus Astronomical Center, Polish Academy of Sciences, Bartycka 18, PL-00-716 Warszawa, Poland\\
 \email{nicolas.scepi@gmail.com}}

   \date{Received September 15, 1996; accepted March 16, 1997}


  \abstract
   {X-ray binaries (XRBs) exhibit spectral hysteresis for luminosities in the range $10^{-2}\lesssim L/L_\mathrm{Edd}\lesssim 0.3$, with a hard X-ray spectral state that persists from quiescent luminosities up to $\gtrsim 0.3L_\mathrm{Edd}$, transitioning to a soft spectral state that survives with decreasing luminosities down to $\sim 10^{-2}L_\mathrm{Edd}$.}
   {We present a possible approach to explain this behavior based on the thermal properties of a magnetically arrested disk simulation.}
   {By post-processing the simulation to include radiative effects, we solve for all the thermal equilibrium solutions as the accretion rate, $\dot{M}$, varies along the XRB outburst.}
   {For an assumed scaling of the disk scale height and accretion speed with temperature, we find that there exists two solutions in the range of $ 10^{-3}\lesssim\dot{M}/\dot{M}_{\rm Eddington} \lesssim 0.1$ at $r=8\:r_g$ ($ 4\times10^{-2}\lesssim\dot{M}/\dot{M}_{\rm Eddington} \lesssim 0.5$ at $r=3\:r_g$) : a cold, optically thick one and a hot, optically thin one.  This opens the possibility of a natural thermal hysteresis in the right range of luminosities for XRBs. We stress that our scenario for the hysteresis does not require to invoke the strong-ADAF principle nor does it require for the magnetization of the disk to change along the XRB outburst. In fact, our scenario requires a highly magnetized disk in the cold, soft state to reproduce the soft-to-hard state transition at the right luminosities. Hence, a prediction of our scenario is that there should be a jet, although possibly very weakly dissipative, in the soft state of XRBs. We also predict that if active galactic nuclei (AGN) have similar hysteresis cycles and are strongly magnetized, they should undergo a soft-to-hard state transition at much lower $L/L_\mathrm{Edd}$ than XRBs. }
   {}


   \maketitle
%
\defcitealias{scepi2024}{SBD24}
\defcitealias{Shakura}{SS73}

\section{Introduction}

   X-ray binaries (XRBs) are compact binary systems containing a stellar-mass black hole or a neutron star, which accretes from a stellar companion through an accretion disk. A large fraction of XRBs have outburst cycles, during which their bolometric luminosities rise by several orders of magnitudes above the quiescent level \citep{dunn2010,tetarenko2016}. These outbursts last from weeks to years and usually recur on periods of years to decades \citep[see][for a compilation of lightcurves of XRBs]{done2007}. 

One of the most intriguing aspects of XRBs is that they change spectral state during their outburst, with the hardness ratio (ratio of hard to soft X-rays) forming a hysteresis cycle \citep{dunn2010}. For simplicity, we define here two states: the soft state and the hard state \citep{remillard2006}. Schematically, XRBs start in quiescence in a hard state in which they remain up to luminosities of $\gtrsim 30\%$ of Eddington before switching to the soft state (see however failed-transition outbursts: \citealt{alabarta2021}). On their way back to quiescence, XRBs stay in the soft state down to luminosities of $\sim 1\%$ of Eddington before switching back to a hard state and down to quiescence \citep{maccarone2003}. The hysteresis appears between $1\%$ and $\gtrsim 30\%$ Eddington, where XRBs can be found in two different spectral states depending on whether they are on their way up or down in luminosity \citep{dunn2010}. 

The soft state refers to a state dominated by thermal emission peaking around 1 keV, while the hard state refers to a power-law dominated state with hard to inverted photon spectral index $\Gamma\sim 2.1$ to 1.6 and a cutoff around $100$ keV. In both states, a sub-dominant high-energy power-law tail extending to the MeV can be observed \citep{grove1998,laurent2011,jourdain2012,cangemi2023}. The usual interpretation of the soft state is that it originates from a relatively cold ($\approx10^{7}$ K), optically thick disk radiating as a multi-temperature black body (\citealt{Shakura}, hereafter SS73). Since the SS73 model is valid in a very large regime of accretion rate extending from almost Eddington sources to very sub-Eddington sources, it is unclear why XRBs below $1\%$ Eddington are never observed to be in a soft state (see \citealt{shaw2016} for the few known soft-to-hard state transitions at lower luminosities). The usual interpretation of the hard state is that it originates from hot ($\approx 10^{9}$ K), optically thin gas in which hot electrons scatter low energy photons up to high energies by the Inverse Compton process and produce a power-law spectrum via multiple scattering \citep{sunyaev1980}. Finally, the origin of the high-energy power-law tail extending to the MeV is more debated but is usually thought to involve non-thermal electrons accelerated in a jet \citep{zdziarski2012} or a hot region around the black hole such as the plunging region \citep{hankla2022}.

Although most models agree that the hard X-ray emission peaking at 100 keV originates from a hot, optically thin component, called a hot corona (although see \citealt{sridhar2021,groselj2024} for simulations showing the formation of a cold corona), the physical nature of this hot corona is still under debate. At low luminosity, where the electrons and protons are decoupled, it is now accepted that the hot corona can be associated with the entire accretion flow, i.e. a hot, two-temperature, geometrically thick flow \citep{narayan1996,esin1997,petrucci2010,poutanen2014,marcel2018a,dexter2021}. However, at high luminosities analytical calculations and numerical simulations both predict that a hot accretion flow should thermally collapse when the electrons become well-coupled to the protons \citep{yuan2014,dexter2021}, leading to speculation regarding the nature and geometry of the hot X-ray corona at high luminosities. Two different geometries have been debated in the community: 1) a vertically extended geometry, such as a ``lamppost'' model, where the corona is a compact source above the black hole \citep{matt1991,martocchia1996,henri1997,dauser2013} that could be associated with an X-ray emitting jet \citep{markoff2005}, or 2) a horizontally extended geometry where the hot corona is replacing/truncating the inner geometrically thin, optically thick disk in the innermost parts of XRBs \citep{haardt1994,ferreira2006,schnittman2013,marcel2018a,kinch2021}. However, recent measurements of the X-ray polarization angle by the {\it IXPE} mission (Imaging X-ray Polarimetry Explorer) have mostly settled this debate since they have shown that the polarization angle is perpendicular to the disk. This strongly favors models where the corona is along the disk plane \citep{krawczynski2022}.\footnote{Although note that the high X-ray polarization degree seen in Cyg X-1 seems to point towards a moderately relativistic outflowing corona \citep{poutanen2023,dexter2024}.}

To form a horizontal corona that effectively replaces the inner optically thick disk in XRBs is not a trivial task at high luminosities. Indeed, since the density in a hot accretion flow increases with luminosity, we would expect a hot accretion flow to thermally collapse around $L \approx 10^{-2}$ $L_\mathrm{Edd}$ \citep{esin1997,yuan2014}. The two main scenarios to maintain a hot inner flow at high densities rely on either the evaporation of the thin disk by a hot atmosphere \citep{meyer1994,spruit2002} or the presence of a highly magnetized disk \citep{ferreira2006,petrucci2010,oda2012,cao2016,marcel2018a,marcel2018b,marcel2019}. The evaporation scenario has recently been disfavored by shearing-box simulations \citep{bambic2024} while the high magnetization scenario has been highly successful in explaining the phenomenology of XRBs \citep{marcel2018a,marcel2018b}. We will focus on the magnetic scenario in the rest of this paper. 

The main idea behind all strongly magnetized models (see \autoref{sec:dynamical} for a quantitative definition of highly magnetized disks) is that strong magnetic fields enhance the radial speed of the accretion flow and its density scale height \citep{ferreira1995,ferreira2006,oda2012,scepi2024}. This is true in self-similar solutions for the Jet-Emitting-Disk (JED; \citealt{ferreira1995}) and in general relativistic magneto-hydrodynamic (GRMHD) simulations of Magnetically Arrested Disks (MAD; \citealt[][hereafter referred to as \citetalias{scepi2024}]{scepi2024}). As a consequence, for a given accretion rate the density is lower in a strongly magnetized disk than in a weakly magnetized disk, pushing hot, optically thin solutions to higher luminosities. This opens up the possibility of forming a hot, hard X-ray corona at high luminosities in XRBs \citep{ferreira2006,petrucci2010,oda2012,cao2016,marcel2018a,liska2022}. 

Moreover, the magnetization provides a second parameter, with the accretion rate, to produce a hysteresis cycle. Indeed, several authors have suggested that the transitions between the XRB spectral states are caused by sudden changes in the magnetization of the accretion flow \citep{ferreira2006,petrucci2008,igumenshchev2009,begelman2014,cao2016,marcel2019}. When the accretion flow is highly magnetized it forms a hot corona (a hard spectral state) and when the accretion flow is weakly magnetized it forms a cold optically thick disk (a soft spectral state). This idea is supported by the fact that the compact radio emission that is observed in the hard state, which is usually associated with the presence of a magnetic jet, is quenched during the transition to the soft state (\citealt{fender2004,fender2009,corbel2012}: see however \citealt{drappeau2017,peault2019}). However, we still lack an understanding of the mechanisms that could drive these magnetization changes. Most notably, there is no understanding of why the soft-to-hard transition happens at almost the same luminosity in all XRBs if it is driven by a change in magnetization.

Following another approach, \cite{marcel2018a} has investigated a scenario in which the hysteresis of XRBs could be a thermal hysteresis, not a magnetic one. In their scenario, a high-magnetization disk is present all along the hysteresis cycle so that there is no need for unknown mechanisms driving the magnetization changes. However, the authors find that thermal hysteresis can only be formed in strongly magnetized disks in the range $10^{-4}\lesssim L/L_\mathrm{Edd}\lesssim 10^{-2}$. This is at least an order of magnitude too low to explain the phenomenology of XRBs. Recently, \citetalias{scepi2024} have shown that GRMHD simulations of  thin MAD disks share a lot of properties with JED semi-analytical solutions but do differ in several respects, namely the presence of magnetic support in the vertical direction (however see \cite{zimniak2024} (submitted)) and the higher radiative efficiency of the disk. These properties of MADs might be able to push the thermal hysteresis to higher luminosities and motivate us to revisit the scenario of \cite{marcel2018a} in the present paper.

Our paper is structured as follows. In \autoref{sec:dynamical}, we summarize the dynamical results of the GRMHD MAD simulation from \citetalias{scepi2024}. In \autoref{sec:hard_state}, we post-process the simulation of \citetalias{scepi2024} to include the cooling  mechanisms and investigate the thermal structure of the hard state. In  \autoref{sec:soft_state}, we extrapolate the thermal solutions of the thin MAD simulation to the soft state, discuss the presence of thermal hysteresis and compare it to the hysteresis found in JED semi-analytical solutions. At the end of \autoref{sec:soft_state}, we also extrapolate our scenario to the case of active galactic nuclei (AGN) and discuss the possible presence of a jet in the soft state of XRBs. 

\section{Dynamical properties of strongly magnetized disks}\label{sec:dynamical}

\begin{figure}
\includegraphics[width=90mm]{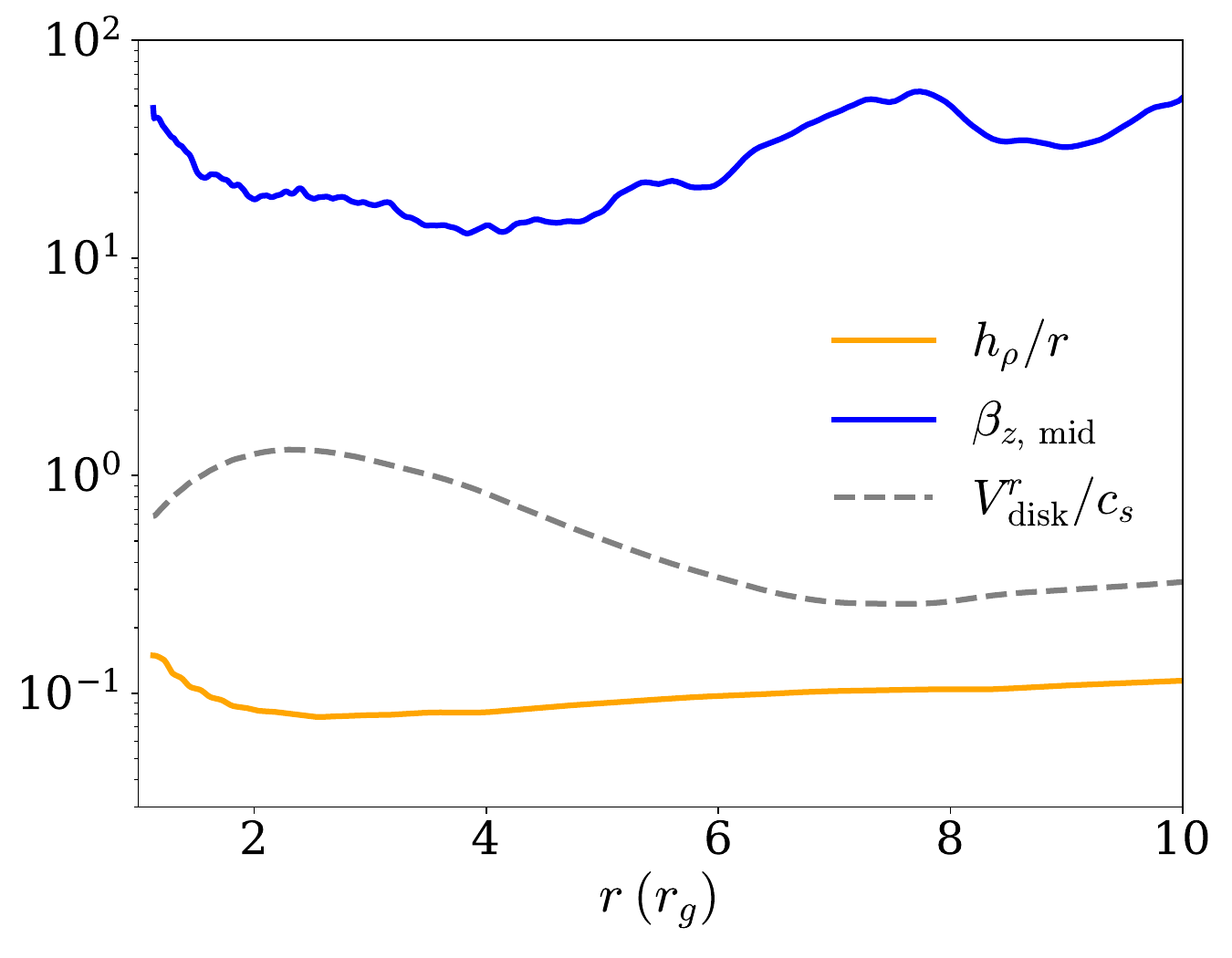}
\caption{The solid gold line shows the geometrical density scale height as a function of radius. The solid blue line shows $\beta_{z,\:\mathrm{mid}}$ at the midplane as a function of radius. The grey dashed line shows the density-weighted radial 3-vector velocity normalized by the sound speed as a function of radius. The density scale-height and the radial speed are much larger than expected in standard theory.}
\label{fig:h_r_ur}
\end{figure}

\begin{figure}
\includegraphics[width=80mm]{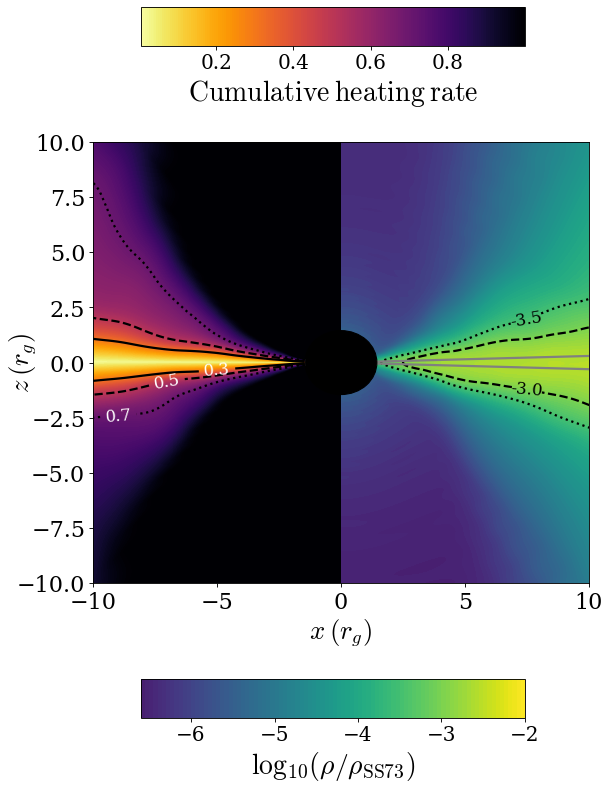}
\caption{Right panel: Time- and $\phi$-averaged log of the density in the simulation of \citetalias{scepi2024} normalized by the density in the midplane that would be obtained from the SS73 model with a same accretion rate, with $\alpha=1$ and $h_\mathrm{th}/r=0.03$. Black lines show three contours of the colormap. Grey lines show the surfaces of $z/r=\pm0.03$. Left panel:  Time- and $\phi$-averaged cumulative heating rate, i.e. the ratio of the heating rate latitudinally integrated up to $\theta$ to heating rate latitudinally integrated up to $\infty$. Black lines show three contours of the cumulative heating rate. }
\label{fig:rho_heating}
\end{figure}

 We refer to strongly magnetized disks as disks having a magnetic field of dynamical importance, or more quantitatively, as disks having a large-scale vertical field, $B_z$ such that $\beta_{z,\:\mathrm{mid}}\equiv 2 (p_\mathrm{gas}+p_\mathrm{rad})/B^2_{z,\mathrm{lam}}(z=0)$ is $\lesssim100$. Here $p_\mathrm{gas}$, $p_\mathrm{rad}$ and $B_{z,\mathrm{lam}}$ are respectively the gas pressure, the radiation pressure and the laminar vertical field. Note that, although a vertical field with $\beta_{z,\:\mathrm{mid}}\approx100$ is not dynamically important by itself, it triggers the establishment of other field components (both turbulent and laminar) through the action of the magneto-rotational instability and of magnetic winds that are dynamically important. Hence, we stress the importance of defining strongly magnetized disks with the $\beta_{z,\:\mathrm{mid}}$ parameter and not the $\beta$ parameter including all the other components \citep{jacquemin2021}. Indeed, because of the indestructibility of net magnetic flux, $B_z$ can be regarded as a control parameter of the system whereas the other field components are an outcome of physical mechanisms depending on $B_z$. Note that our definition of strongly magnetized disks includes both JED semi-analytical solutions and MAD numerical simulations, which share many dynamical properties \citep{scepi2024}. \newline

In this paper, we post-process the results of the GRMHD simulation of a thin, MAD disk by \citetalias{scepi2024}. The simulation was performed with the code {\sc Athena++} \citep{white2016,stone2020} using Kerr-Schild coordinates and was initialized with a \cite{fishbone1976} torus with an inner radius of $16.45\:r_g$ and a pressure maximum at $34\:r_g$, where $r_g\equiv GM_\mathrm{BH}/c^2$ is the gravitational radius, $G$ is the gravitational constant, $M_\mathrm{BH}$ the mass of the black hole and $c$ the speed of light. The black hole spin is $a=0.9375$ and the $\beta_{z,\:\mathrm{mid}}$ parameter is $\approx100$ throughout the initial torus. The effective resolution is $1024\times 512\times 1024$ (four levels of static mesh refinement were used) for a domain of size $[1.125\:r_g,1500\:r_g],[0,\pi],[0,2\pi]$ in the $r,\theta$ and $\phi$ directions,  respectively. A snapshot of the simulation at $40,400\:r_g/c$ is used for the analysis in the present paper. We also use time-averaged and $\phi-$averaged data from \citetalias{scepi2024} for the heating rate\footnote{See \citetalias{scepi2024} for the definition of the heating rate.} in \autoref{sec:hard_state} and for all quantities in \autoref{sec:soft_state}. The reason why we always use a time and $\phi-$ averaged version of the heating rate is that it is a noisy quantity. Time-averages are done between $40,000$ and $41,000$ $r_g/c$ to isolate a local moment in time around our snapshot while having good statistics on the heating rate. We also tried the same averaging window as in \citetalias{scepi2024}, i.e. between $40,000$ and $44,000$ $r_g/c$, and do not find any substantial difference in our results. 

To obtain a thin disk, \citetalias{scepi2024} used an artificial cooling function to cool the disk to a targeted thermal scale height of $h_\mathrm{th}/r=0.03$, where 
\begin{equation}
    \frac{h_\mathrm{th}}{r}\equiv \sqrt{\frac{2}{\pi}}\frac{c_\mathrm{s}}{v_\mathrm{K}},
\end{equation}
 $c_s$ is the sound speed at the midplane and $v_K$ is the relativistic Keplerian velocity around a Kerr black hole defined as in \cite{noble2009}. The disk eventually became a thin MAD  \citep{igumenshchev2008}, with $\beta_{z,\:\mathrm{mid}} \approx 26$ when averaged over $r<10\:r_g$ (see \autoref{fig:h_r_ur}) and a $\beta$ parameter (also at the midplane but including all field components) $\approx0.2$, so that magnetic effects dominate the dynamics of the disk. 

The dominance of the magnetic field makes the disk dynamical properties diverge from standard theory in the following respects:
\begin{enumerate}
    \item the vertical structure is supported by the gradient of turbulent magnetic pressure in the disk, not by the gradient of thermal pressure as in SS73. Indeed, we plot on the top panel of \autoref{fig:h_r_ur} the geometrical scale height of the disk,
    \begin{equation}
    h_\rho\equiv \frac{\int_\theta \sqrt{-g} \rho|\theta-\theta_0| d\theta}{\int_\theta \sqrt{-g} \rho d\theta}
    \end{equation}
    where $\theta_0\equiv \frac{\pi}{2} + \int \rho (\theta-\pi/2) d\theta/\int \rho d\theta$ as in \cite{mckinney2012}. We find that $h_\rho/r\approx 0.09$ for $h_\mathrm{th}/r=0.03$, so that the geometrical scale height is three times larger than the thermal scale height. 
    \item the accretion is mostly due to large-scale magnetic stress originating from a magnetized wind \citep{blandford1982,ferreira1993}, not to viscous stress as in SS73. As a result, the accretion speed is much higher than in standard theory, as is also found in JED solutions \citep{ferreira1995,ferreira1997}. This can be seen on the bottom panel of \autoref{fig:h_r_ur}, where we plot the accretion speed normalized by the sound speed, $v_r/c_s\equiv \int_\theta \sqrt{-g}\rho V^r d\theta / \int_\theta \sqrt{-g}\sqrt{\mathrm{\rho} p_\mathrm{gas}} d\theta $, where $V$ is the three-vector velocity. We find $v_r/c_s\approx 0.8$ when averaged over $r<10\:r_g$, which can be compared with a prediction of $3\times10^{-2}$ for $\alpha=1$ in standard theory. 
    \item the dissipation profile does not follow the thermal pressure latitudinal profile as in SS73 as it shows significant dissipation at the base of the wind. This can be seen in \autoref{fig:rho_heating} where we show contours of the ratio of the cumulative heating rate in the latitudinal direction up to $\theta$ to the cumulative heating rate in the latitudinal direction up to $\infty$. Less than $30\%$ of the dissipation happens within $z/r\pm 0.03$ (i.e $z/h_\mathrm{th}\pm1$). 
    \item there is a significant amount of heating inside the innermost stable circular orbit (ISCO) \citep{agol2000}, which provides additional dissipation compared to the SS73 model. 
\end{enumerate}

An important consequence of magnetic support and wind-driven accretion is that the disk is much less dense than expected in standard theory \citep{ferreira2006}. This can be seen in \autoref{fig:rho_heating} where the right panel shows the density in the \citetalias{scepi2024} simulation, divided by the expected density in standard theory if the disk were accreting at the same accretion rate with $\alpha=1$. We see that, even in the midplane, the disk density is three orders of magnitude lower than expected. We also see that the density slope in the MAD simulation is very shallow, scaling as $\propto r^{-1}$ compared to $\propto r^{-15/8}$ in standard theory. Note that this is actually the slope expected in a JED with an ejection index of $\approx 0.5$ such as in \citetalias{scepi2024}, showing once again the similarity between JED semi-analytical solutions and MAD numerical simulations. Finally, we stress that it is the low densities of strongly magnetized disks, coupled to the non-zero magnetic stress at the ISCO in the MAD simulation, that will enable the formation of a hot coronal state at high accretion rates, as we will see in the next section.

\section{Post-processing of the hard state}\label{sec:hard_state}
\subsection{Post-processing procedure}\label{sec:thermal_prop}
 In this section, we post-process the simulation of \citetalias{scepi2024} to solve for the gas temperature.  We will assume that ions and electrons are well-coupled so that the plasma has only one temperature. We will discuss this assumption later along with \autoref{fig:yc_hard}. We will also neglect feedback of thermal pressure and radiation pressure on the dynamical structure of the disk. This choice is motivated by the dominance of magnetic effects over thermal effects in setting the dynamical structure of the strongly magnetized disk in \citetalias{scepi2024}. This neglect of thermal feedback is likely to hold as long as the final post-processed temperature does not depart too much from the targeted temperature used in \citetalias{scepi2024} (see \autoref{sec:soft_state} for when we impose large deviations). As the targeted thermal scale height of $h_\mathrm{th}/r=0.03$ used in \citetalias{scepi2024} sets a temperature between $10^8$ and $10^9\:K$ in the inner $10\:r_g$ of an X-ray binary, this makes our post-processing particularly relevant to the study of the hot coronal, hard spectral state. Note, however, that in the cold solutions displayed in \autoref{fig:hysteresis}, where the departure from the targeted thermal scale height of $h_\mathrm{th}/r=0.03$ is large, the radiation pressure starts to dominate over the magnetic pressure for $\dot{m}>10^{-2}$ so that the structure of the disk should be altered by the radiative effects in this case. Taking those effects into account is beyond the scope of this paper.

 To solve for the gas temperature, we equate the fluid-frame heating rate per unit of proper time and volume as defined and measured in \citetalias{scepi2024}, $q_\mathrm{heat}$, with a cooling rate that we estimate in post-processing, $q_\mathrm{cool}$, following \cite{esin1996}. Note that the heating rate is actually a measure of the amount of internal energy that is removed to keep a constant temperature in the disk in the simulation of \citetalias{scepi2024}. Hence, it also takes into account heat advection. 
 
To compute the cooling rates, we make a few assumptions. We neglect bound-free and bound-bound cooling processes as the plasma in XRBs is expected to be highly ionized. We also neglect external Compton cooling, which is found to be sub-dominant at high luminosities ($\approx 10\% L_\mathrm{Edd}$) in the JED solution of \cite{marcel2018b}\footnote{Note that the importance of external Compton also depends a lot on the geometry of the outer optically thick disk that irradiates the inner optically thin corona. For a very flared outer disk, external Compton could become important (see \cite{marcel2018b}).}. This means that we only consider synchrotron self-Compton, denoted as $q_\mathrm{synch,C}$, and bremsstrahlung self-Compton emission, denoted as $q_\mathrm{brem,C}$, with the formulas taken from \citealt{esin1996}. Hence, the cooling rate in the optically thin regime, which is purely local, is given by 
 \begin{equation}
     q_\mathrm{thin} = q_\mathrm{synch,C} + q_\mathrm{brem,C}.
 \end{equation}
 At large optical depths, we also take into account black-body emission given by
 \begin{equation}
     q_\mathrm{thick} = \frac{4\sigma_B T_e^4}{3\tau_\mathrm{tot} h_\rho},
 \end{equation}
 where $T_e$ is the electron temperature. Here $\tau_\mathrm{tot}$ is a local estimate of the total optical depth
 \begin{equation}
     \tau_\mathrm{tot} = \tau_R + \tau_\mathrm{es} \equiv (\kappa_\mathrm{R}+\kappa_\mathrm{es})\rho h_\rho,
 \end{equation}
 where $\kappa_\mathrm{R}=5\times10^{24} \rho T^{-3.5} \mathrm{g\:cm^{-2}}$ is the Rosseland mean opacity and $\kappa_\mathrm{es}$ is the electron scattering opacity. \footnote{Note that our computation of opacities can only be an approximation without having proper radiative transfer following the path of individual photons}. 
 Finally, we make the transition from the optically thin to optically thick regime by using the bridge formula from \citealt{artemova1996},
 \begin{equation}\label{eq:bridge}
     q_\mathrm{cool} = q_\mathrm{thick}\left(1+\frac{4}{3 \tau_\mathrm{tot}}\left[1+\frac{1}{2\tau_\mathrm{abs}}\right]\right)^{-1}.
 \end{equation}
 where we have defined the optical depth to absorption as
 \begin{equation}
     \tau_\mathrm{abs} = \frac{2 q_\mathrm{thin}}{3 q_\mathrm{thick} \tau_\mathrm{tot}}.
 \end{equation}
One can easily verify that in the effectively optically thin regime, when $\tau_\mathrm{abs}\ll 1$ and $\sqrt{\tau_\mathrm{tot}\tau_\mathrm{abs}}\ll 1$, \autoref{eq:bridge} tends to $q_\mathrm{thin}$ while in the optically thick regime, when $\tau_\mathrm{abs}\gg 1$ and $\tau_\mathrm{tot}\gg 1$ , \autoref{eq:bridge} tends to $q_\mathrm{thick}$ \citep{artemova1996}.

 The density in the ideal MHD simulation of \citetalias{scepi2024} is a dimensionless parameter, which we convert to physical units using
 \begin{equation}\label{eq:rho_cgs}
     \rho_\mathrm{cgs} = \rho_\mathrm{code}\frac{M_\mathrm{unit}}{V_\mathrm{unit}},
 \end{equation}
 where $M_\mathrm{unit}=\dot{M}_\mathrm{XRB}/\dot{M}_\mathrm{code}\times t_\mathrm{unit}$, $V_\mathrm{unit}\equiv (G M_\mathrm{BH}/c^2)^3$ and $t_\mathrm{unit}\equiv G M_\mathrm{BH}/c^3$. Here $\dot{M}_\mathrm{XRB}$, $\dot{M}_\mathrm{code}$ and $M_\mathrm{BH}$ are respectively the accretion rate in cgs units, the accretion rate in code units from the simulation of \citetalias{scepi2024} and the mass of the central black hole in cgs units. In the following, we use $M_\mathrm{XRB}=10 M_\odot$ and $\dot{m}= \dot{M}_\mathrm{XRB}/\dot{M}_\mathrm{Edd}$, with $\dot{M}_\mathrm{Edd}c^2\equiv L_\mathrm{Edd}/0.1$, to set the accretion rate.

 To summarize, in order to find the temperature in post-processing we solve the following equation
\begin{equation}\label{eq:heating_equal_cooling}
q_\mathrm{heat}(\rho) = q_\mathrm{cool}(\rho, T, B),
\end{equation}
where the choice of a black hole mass and an accretion rate sets the physical units of $\rho$ and of the magnetic field, $B$.
  
\subsection{Temperatures in the hard state}

\begin{figure*}
\includegraphics[width=0.92\textwidth]{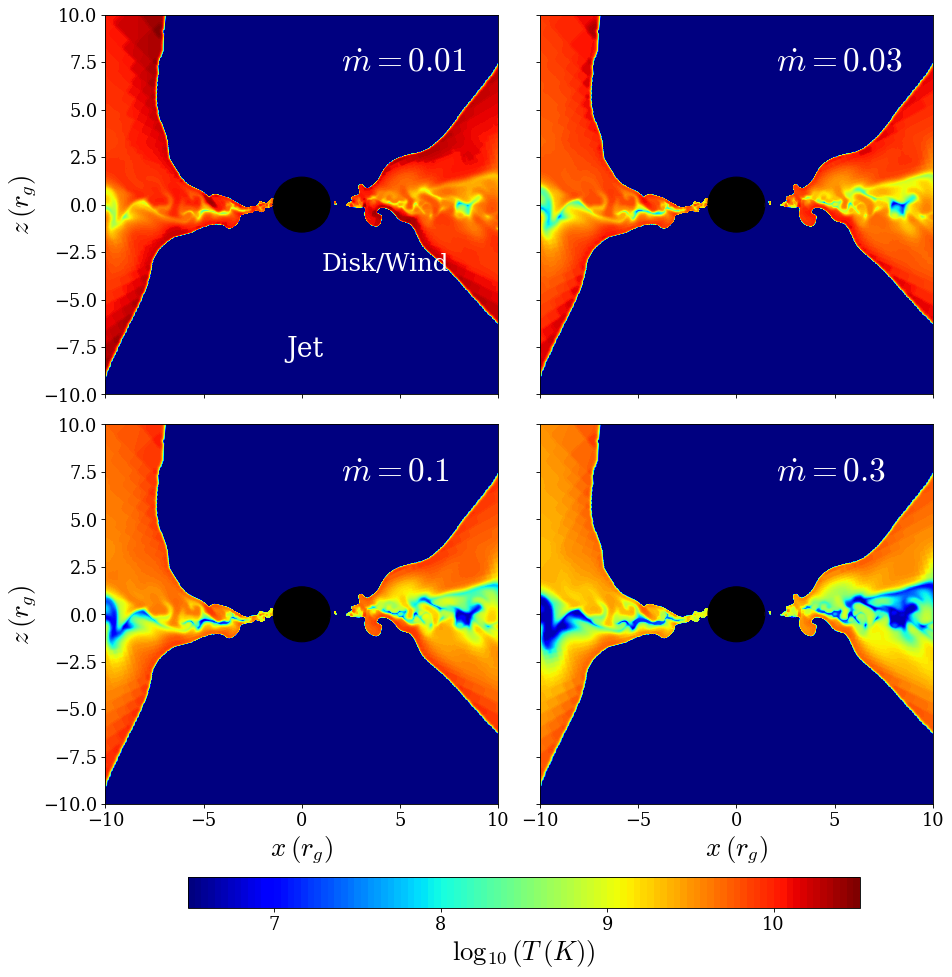}
\caption{Poloidal cuts of the post-processed temperature for four different accretion rates. The jet region (defined as the region where $\sigma>1$) is excluded from our analysis.}
\label{fig:poloidal_T_hard}
\end{figure*}

\begin{figure*}
\includegraphics[width=0.92\textwidth]{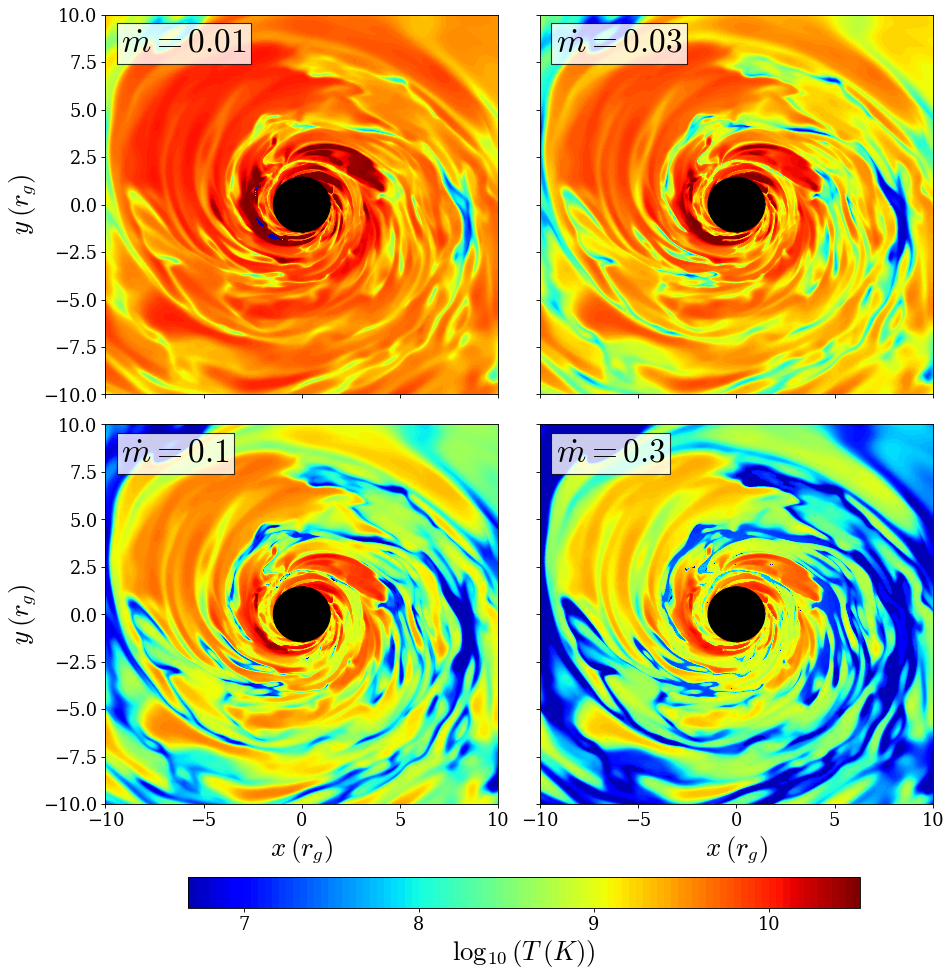}
\caption{Midplane cut of the post-processed temperature for four different accretion rates. The temperature is inhomogeneous but even at high accretion rates, patches of hot gas can survive.}
\label{fig:midplane_T_hard}
\end{figure*}

\autoref{fig:poloidal_T_hard} and \autoref{fig:midplane_T_hard} show a poloidal cut and a midplane cut, respectively, of the post-processed temperature for different accretion rates. In the poloidal cut, we exclude the jet region (defined here as where the cold magnetization $\sigma\equiv B^2/\rho$ is larger than unity) from our analysis. 

In all cases, we find that a significant fraction of the gas can stay hot, even at high accretion rates. For example, for $\dot{m}=0.3$, $\approx 75\%$ of the heat is dissipated in regions that have temperatures above $10^8$ K. In fact, the averaged temperature weighted by $q_\mathrm{heat}$ goes from $\approx 10^{10}$ K for $\dot{m}=0.01$ to $\approx 10^{9}$ K for  $\dot{m}=0.3$. We note that the high temperature region in the left corner of the panels of \autoref{fig:midplane_T_hard} is related to the presence of a past MAD eruption (see \cite{igumenshchev2009,mckinney2012} for literature on the MAD eruptions). This region is of relatively low-density, high magnetization and is typically dominated by synchrotron self-Compton emission (see  \autoref{fig:cooling_midplane}).  Remnants of MAD eruptions are present at any given time within $10\:r_g$ so that these eruptions might play a large role in producing high temperature spots in the inner disk. 

\begin{figure*}
\includegraphics[width=0.8\textwidth]{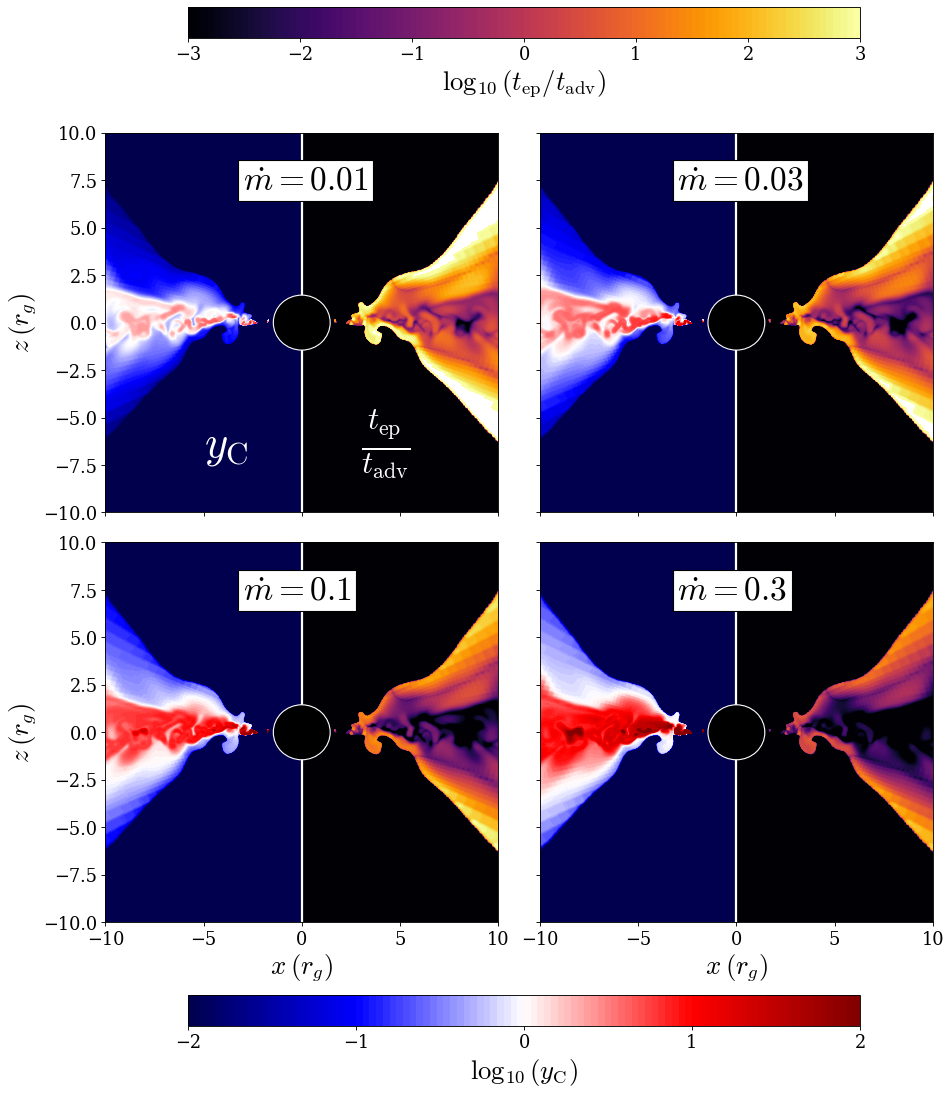}
\caption{Poloidal cut of the $y_\mathrm{C}$ Compton parameter and the ratio of electron-proton collision time to the advection time.}
\label{fig:yc_hard}
\end{figure*}

We also plot on \autoref{fig:yc_hard} the $y_C$ Compton parameter in the non-relativistic limit, 
\begin{equation}
y_\mathrm{C} = \frac{4  k_b T}{m_e c^2}\mathrm{Max}(\tau_\mathrm{es},\tau_\mathrm{es}^2).
\end{equation}
The averaged $y_\mathrm{C}$ weighted by $q_\mathrm{heat}$ goes from $\approx 0.8$ at $\dot{m}=0.01$ to $\approx 20$ for  $\dot{m}=0.3$ so that a relatively hard spectral index could be expected from a spectral post-processing of our simulations. 

We also note the presence of localized dense clumps with temperatures around $\approx 10^{6-7}$ in the body of the disk, which cool down through black-body emission (as can be seen in \autoref{fig:cooling_poloidal} and \autoref{fig:cooling_midplane}). These clumps appear in the outer disk at low accretion rate and spread to the inner disk as the accretion rate increases. This is reminiscent of the clumps seen in \cite{liska2022}. Note that these cold clumps could cool the surrounding hot gas by external Compton, an effect which we neglect here. However, these clumps are also embedded in hot gas and so it remains to be seen whether they could survive when irradiated by the surrounding hard radiation. 

By averaging over time and $\phi$, we can compute the density weighted temperature as a function of radius and define a ``truncation'' radius, i.e. a radius within  which the disk is hotter than $10^8\mathrm{\:K}$. We find that for $\dot{m}=0.01$ it is $> 10\:r_g$ (so outside of our analysis radius) while for $\dot{m}=0.3$ it is $\approx 7\:r_g$. The truncation thus naturally recedes inward as $\dot{m}$ increases. We emphasize that the truncation radius does not recede inwards because of a change in the magnetization of the disk. Indeed, in our normalization procedure we assume that the magnetization stays the same for all $\dot{m}$. The truncation radius simply traces the radius at which the disk cannot stay hot anymore because otherwise the cooling rate would exceed the heating rate. The fact that the truncation radius appears in the outer part and propagates inward follows from the shallowness of the density profile in \citetalias{scepi2024}. As the density is relatively shallow, going as $r^{-1}$, the cooling rate starts dominating over the heating rate in the outer parts first, with bremsstrahlung SC emission and black-body emission becoming increasingly important in the outer parts of the disk while the inner disk is mostly cooled through synchrotron SC (see \autoref{fig:cooling_poloidal} and \autoref{fig:cooling_midplane}). This result is also found in the JED solutions of \cite{marcel2018a} but is in contrast with the hot ADAF (advection-dominated accretion flow) solutions where the cold disk solution appears in the inner disk as $\dot{m}$ increases because of the steepness of the density profile in the ADAF solution \citep{esin1996}.

Finally, we verify our hypothesis that the fluid is one-temperature by plotting on \autoref{fig:yc_hard} the electron-proton collision time, $t_\mathrm{ep}$, divided by the inflow/outflow time, $t_\mathrm{adv}$. We define $t_\mathrm{ep}$ as in \cite{mahadevan1997}, where we assume the same temperature for electrons and protons so as to give an upper estimate, and $t_\mathrm{adv}=r/|V^r|$ where $V^r$ is the time- and $\phi$-averaged 3-vector velocity from \citetalias{scepi2024}. We find that  the averaged $t_\mathrm{ep}/t_\mathrm{adv}$ weighted by $q_\mathrm{heat}$ is $\approx 1$ for $\dot{m}=0.01$ while it is $\approx10^{-2}$ for $\dot{m}=0.3$. This means that, while the bulk of the disk might be able to have different proton and electron temperatures at $\dot{m}=0.01$, the disk should be one-temperature for  $\dot{m}>0.01$. Note that this result is at odds with \cite{liska2022}, which finds that a thin MAD can remain two-temperature at luminosities as high as $L=0.3\:L_\mathrm{Edd}$, but is consistent with \cite{dexter2021}, which finds that electrons and protons in a MAD become well-coupled for $\dot{m}>0.01$. 

\section{Exploration of the soft state and thermal hysteresis}\label{sec:soft_state}

\begin{figure*}
\includegraphics[width=0.8\textwidth]{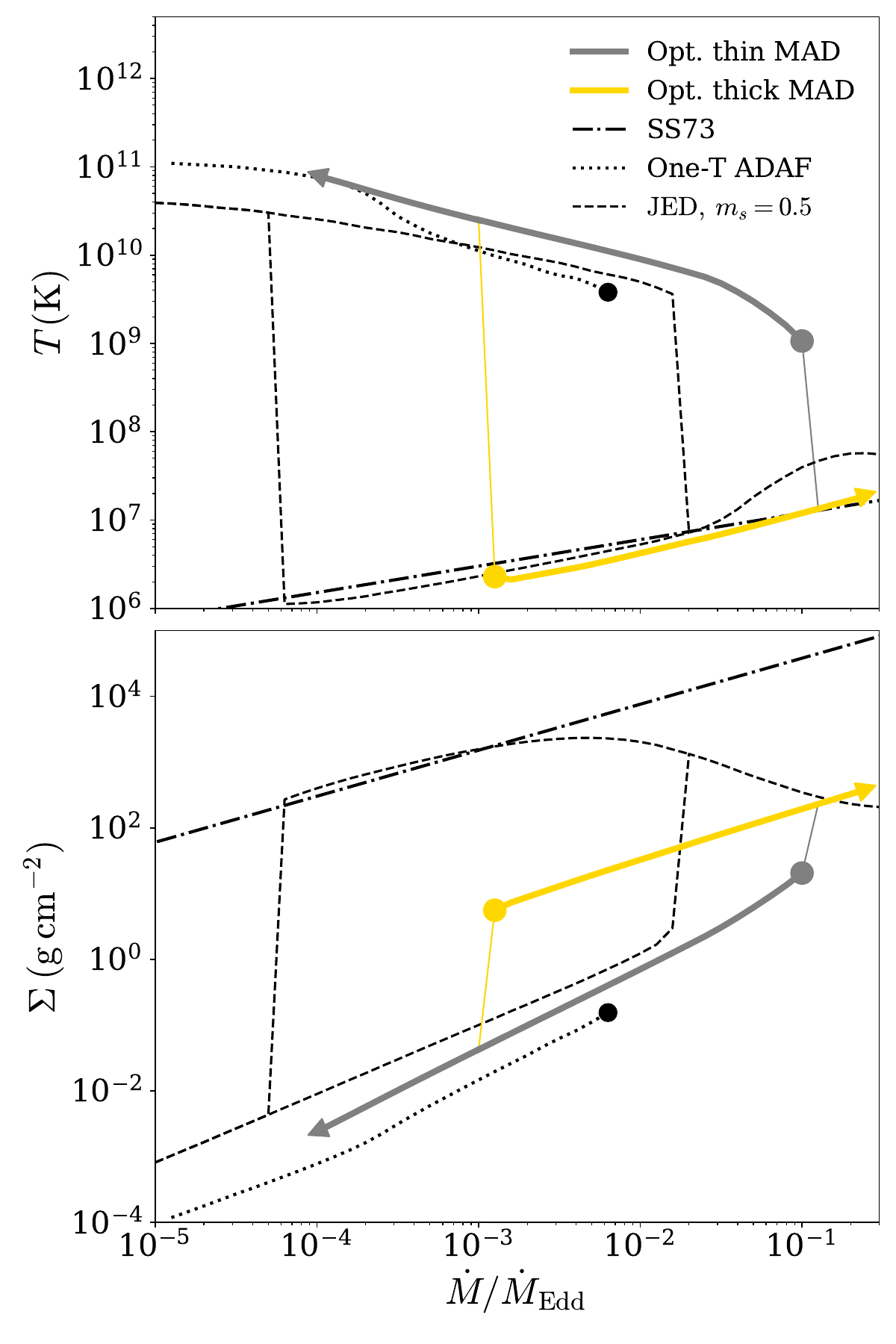}
\caption{Thermal solutions of the MAD numerical simulation at $r=8\:r_g$ for an XRB with $M_{\rm BH}=10 M_\odot$. In both panels, the grey line shows the hot, optically thin branch that extends up to $\dot{m}\approx10^{-1}$ while the gold line shows the cold, optically thick branch that only extends down to $\dot{m}\approx10^{-3}$. The black lines show reference solutions for comparison with the dashdotted, dotted and dashed lines showing the solutions for, respectively, a standard Shakura-Sunyaev disk, a one-temperature ADAF analytical solution and a two-temperature JED semi-analytical solution. The uniqueness of our MAD thermal solutions is that they have two stable thermal branches within the range $10^{-3}\lesssim \dot{m}\lesssim 10^{-1}$.}
\label{fig:hysteresis}
\end{figure*}

\begin{figure}
\includegraphics[width=80mm]{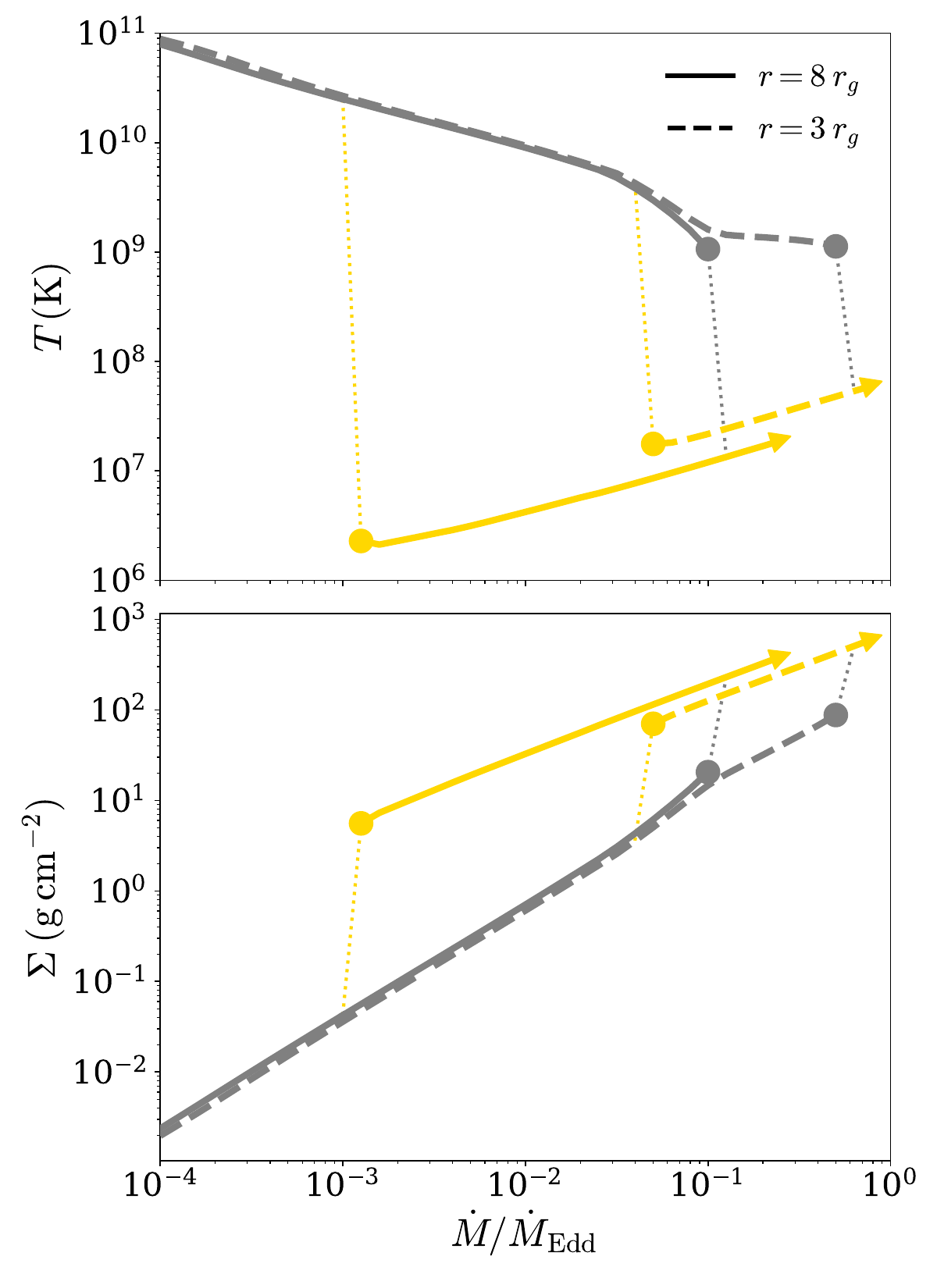}
\caption{Thermal solutions of the MAD numerical simulation for an XRB with $M_{\rm BH}=10 M_\odot$ at two different radii of $r=8\:r_g$ (solid lines) and $r=3\:r_g$ (dashed lines). The grey lines show the hot, optically thin branches and the gold lines show the cold, optically thick branches. The inner radii can stay hot to higher $\dot{m}$ than the outer radii. }
\label{fig:XRB_two_r}
\end{figure}

\subsection{Thermal solutions of strongly magnetized disks}\label{sec:thermal_curves}

In this subsection, we compute the thermal structure of the disk along an entire XRB outburst. The main difference with the preceding section is that we solve for the $\theta$-averaged (along with time and $\phi$-averaged as before) thermal structure of the disk. Also, to extrapolate the behavior of strongly magnetized disks to the very thin and cold disk regime, we make a few additional physically motivated assumptions regarding how the disk properties evolve with $h_\mathrm{th}/r$, i.e with the temperature. 
Indeed, the simulation of \citetalias{scepi2024} has $h_\mathrm{th}/r=0.03$, which roughly gives a temperature of $\approx 4\times 10^8$ K at $r=10 r_g$ in an XRB. This is well suited for a study of the hard state but is very hot compared to the soft state where we can expect $h_\mathrm{th}/r$ to be as low as $3\times 10^{-3}$.

First, we assume that although the effective scale height of the disk is larger (by a factor of 3 in \citetalias{scepi2024}) than the thermal scale height, the former will scale proportionally to the latter such that
\begin{equation}
    h_\rho \approx 3 h_\mathrm{th} \propto \sqrt{T}.
\end{equation}
In the same spirit, we assume that the accretion speed will scale proportionally to the sound speed such that
\begin{equation}
    v_r \propto c_s \propto \sqrt{T}.
\end{equation}
Note that this scaling is equivalent to $v_r\propto (h_\mathrm{th}/r)$, which is the expected scaling when angular momentum is removed mostly through large-scale wind-driven torques \citep{ferreira1995}.

These assumptions are motivated by our working hypothesis that the magnetization stays the same throughout an outburst (thanks to a rapid reorganization of the magnetic field as we discuss in \autoref{sec:hysteresis}). In reality, the actual ratios between the disk scale height and the thermal scale height as well as the accretion speed and the sound speed are likely to be functions of the magnetization, which we assume here to be constant for simplicity. We emphasize that these two assumptions are crucial ingredients of our scenario, especially for the existence of a cold, optically thick solution as we will see later on in this section. 

Put together, these two assumptions give the following scaling for the $\theta$-averaged density in the disk,
\begin{equation}
    \rho_0 = \frac{\dot{M}}{2\pi r v_r 2h_\rho} \propto {T}^{-1}.
\end{equation}

Hence, to compute the thermal solutions of strongly magnetized disks, we use the scaling of $\rho_0$ with the temperature to renormalize the time- $\phi$- \emph{and} $\theta$- averaged run of \citetalias{scepi2024}. This means that we solve for all the different temperatures satisfying \autoref{eq:heating_equal_cooling} at a given radius in the disk, where the $\theta$-averaged density is now given by 
\begin{equation}
     \rho_\mathrm{0,\:cgs,\:renorm}(r) = \rho_\mathrm{0,\:cgs}(r) \frac{T_\mathrm{0.03}(r)}{T(r)},
\end{equation}
where $\rho_\mathrm{0,\:cgs}$ is given by \autoref{eq:rho_cgs} and $T_\mathrm{0.03}(r)$ is the effective temperature of a disk with $h_\mathrm{th}/r=0.03$. 

We also note that other quantities such as the amount of advected energy in the disk, the amount of energy that goes into the wind or the ratio of the turbulent to wind-driven torque are expected to depend on $h_\mathrm{th}/r$. This would introduce an additional dependency of $q_\mathrm{heat}$ and $\rho$ with $h_\mathrm{th}/r$. Given that these dependencies are largely unconstrained in MAD simulations and given the exploratory nature of our study, we will neglect these effects here.

We show in \autoref{fig:hysteresis} the thermal solutions of our strongly magnetized disk at a radius of $8\:r_g$ \footnote{Note that our solutions being one-temperature, the temperature of the optically thin solution for $\dot{m}\lesssim0.01$ is likely to be underestimated. However, since our focus is mostly on the high luminosity regime this approximation is of little consequence here.}. We also show on \autoref{fig:XRB_two_r} the thermal solutions at $3\:r_g$ for comparison. We only show the stable solutions here although there are always three different solutions: two stable ones and one unstable. The grey solid lines show the hot, optically thin branch while the gold solid lines show the cold, optically thick branch. We also show the standard solution of the one-temperature ADAF of \cite{esin1996} as a dotted black line and the standard optically thick disk of \citetalias{Shakura} as a dash-dotted black line. The end of a solution is denoted by a filled circle, while the continuation of a solution is denoted by an arrow.

The most interesting feature of the MAD simulation of \citetalias{scepi2024} is that it can have two stable thermal solutions at the high-end of the $\dot{m}$ range. For $r=8\:r_g$, two solutions exist in the range $0.001 \lesssim \dot{m} \lesssim 0.1$, while for $r=3\:r_g$ two solutions exist in the range $0.04 \lesssim \dot{m} \lesssim 0.5$. The inner disk stays hot to larger $\dot{m}$ and the disk cools from outside-in. For $r=8\:r_g$, we see that the optically thin solution exists up to $\dot{m}\lesssim 0.1$ so that our optically thin strongly magnetized solution can survive to values of $\dot m$ an order of magnitude higher than the one-temperature hot flow solution of \cite{esin1996}. In fact, the optically thin MAD solution acts as a prolongation at high $\dot{m}$ of the solution from \cite{esin1996}. Our MAD simulation also has an optically thick solution that exists down to $\dot{m}\gtrsim 0.001$ for $r=8\:r_g$ and $\dot{m}\gtrsim 0.04$ for $r=3\:r_g$. It is interesting to note that the optically thick MAD solution does not exist at any radius at low accretion rates, in contrast to the \citetalias{Shakura} solution, which is a valid solution down to $\dot{m}$ as low as $\approx10^{-14}$ at which point it becomes optically thin. This is because the surface density is much lower (almost two orders of magnitude) in the MAD numerical solution than in the \citetalias{Shakura} analytical solution. 

We also show in \autoref{fig:hysteresis} the comparison between the thermal solutions coming from the MAD numerical simulation of \citetalias{scepi2024} and a two-temperature JED semi-analytical solution of \cite{marcel2018a} with a sonic Mach number $m_s\equiv v_r/c_s=0.5$ and $\beta_{z,\:\mathrm{mid}}=20$. We see that the JED solution has two thermal solutions in the range $6\times10^{-5}\lesssim \dot{m} \lesssim 2\times 10^{-2}$ but cannot stay optically thin for $\dot{m}\gtrsim 2\times 10^{-2}$. \footnote{To fit the X-ray spectrum of XRBs at high luminosity, such as GX 339-4 \citep{marcel2019} or MAXI J1820+070 \citep{marino2021}, a larger $m_s = 1.5$ is used for the JED semi-analytical solution. However, note that those solutions at high $m_s$ do not posses a cold optically thick thermal solution.} Our MAD simulation has $m_s\approx0.8$ and $\beta_{z,\:\mathrm{mid}}\approx 26$ (see \autoref{fig:h_r_ur}). Despite this $37\%$ difference in $m_s$ (translating in a difference in $\Sigma$), the two situations can be readily compared and we see that the two stable thermal branches are shifted. We see that in the MAD thermal solutions the two stable thermal branches are shifted to higher $\dot{m}$ by a factor of $\approx6$ compared to the JED thermal solutions. We will see in \autoref{sec:comparison} that the reasons behind this difference lie in another phenomenon. 

\subsection{A tentative hysteresis cycle}\label{sec:hysteresis}
There is ample evidence that the outbursts of X-ray binaries are triggered in the outer disk \citep{lasota2001}.  One promising mechanism to trigger the eruption is the ionization instability model (see \cite{coriat2012} for observational evidence supporting this mechanism) where the instability triggers an increase in the accretion rate in the outer parts of the disk, which then propagates inward and puts the entire disk in a highly accreting state. This is the rise to outburst. As accretion proceeds, it depletes the disk until the ionization instability is triggered again where the density is too low. This is the return to quiescence \citep{dubus2001,lasota2001}. In this framework, all the timescales involved in the X-ray binary hysteresis are then governed by the outer disk and we can treat the inner disk as being in a quasi-steady state with a constant $\dot{m}$.

Although the hysteresis is a global problem involving a range of radii, we will explain our scenario using a local approach, placing ourselves at $r=8\:r_g$, for simplicity. We have seen in the previous section that, given our assumption that $h_\rho\propto T^{1/2}$, the MAD solution of \citetalias{scepi2024} has two thermal equilibria in the range $0.001 \lesssim \dot{m} \lesssim 0.1$ at $r=8\:r_g$. We now make the crude approximation that the radiative efficiency is the same for the two thermal solutions above $\dot{m}\gtrsim 0.01$ and is $\approx 0.2$, a value close to what is generally found in MAD simulations \citep{avara2016,dexter2021,liska2022,scepi2024}. With this radiative efficiency, our two solutions have the potential of producing a thermal hysteresis in the range of luminosities $0.004 \lesssim L \lesssim 0.2$. This is very reminiscent of the range at which spectral hysteresis is observed in X-ray binaries. Of course, the radiative efficiency could be lower in the hard state than in the soft state by as much as a factor of 5 (see Fig.6 of \citealt{marcel2018b}). This would reduce the maximum luminosity of the hard state and increase the maximum luminosity of the soft state. However, other effects such as X-ray absorption along the line of sight should be taken into account when comparing in detail with observations and our assumption of a constant radiative efficiency is only a first step for this exploratory paper. 

We have seen in \autoref{sec:hard_state} that the hot, optically thin (cold, optically thick respectively) MAD solution has qualitatively the right temperatures to produce a hard state (soft state\footnote{We note that the relatively low densities and the strong dissipation away from the midplane in our cold, optically thick solution might produce large color corrections. This might make it challenging to reproduce the spectrum of XRBs with very soft states, such as LMC-X3 where measurements of spin through continuum fitting with moderate color corrections already give low value of the spin \citep{davis2006,yilmaz2023}. However, large color correction might help to explain the apparent sizes of AGN \citep{hall2018}.} respectively). Now, we can draw a sketch for a tentative spectral hysteresis cycle. We will assume that the inner disk stays MAD (i.e. stays highly magnetized) all along the outburst and assume a radiative efficiency of $0.2$ for simplicity. The disk starts in quiescence and $\dot{m}$ starts increasing at the beginning of the outburst. It has been shown by \cite{dexter2021} that MADs can reproduce the properties of the hard state (power-law and cutoff evolution) from quiescence to $L\lesssim 10^{-3}\:L_\mathrm{Edd}$. For $L\gtrsim 10^{-3}\:L_\mathrm{Edd}$, the disk will start to cool to temperatures close to $10^{9}$ K \citep{dexter2021}, at which point the magnetization increases even further (\citetalias{scepi2024}). This extremely magnetically dominated state can sustain a hot, optically thin solution, which can be seen as a magnetized extension of the hot one-temperature flow of \cite{esin1996} (see \autoref{fig:hysteresis}), as found also in the JED solutions of \cite{petrucci2010,marcel2018a}. Consequently the disk naturally transits from one optically thin solution to the other. The disk will track the optically thin solution until the latter ceases to exist. In a MAD, this happens at $L\approx 0.2\:L_\mathrm{Edd}$ (at $r=8\:r_g$), although the exact value would change if on-the-fly radiation were included in the simulation. At this point, the disk has to switch to the only available solution, which is the cold, optically thick one. This corresponds to the transition to the soft state at high luminosities.  Similarly, as $\dot{m}$ decreases back to quiescence, the disk will stay on the cold, optically thick solution until this solution ceases to exist. In a MAD, this happens for $L\lesssim 0.004\:L_\mathrm{Edd}$ (at $r=8\:r_g)$, where the only available solution is the hot, optically thin one. Again, the exact value will most likely change when more realistic simulations of this very cold state are  available. Nonetheless, the disappearance of the cold solution offers a physical mechanism for X-ray binaries to transit back from the soft state to the hard state at $L\approx 10^{-2}\:L_\mathrm{Edd}$. Finally, the disk returns to quiescence staying in the only available solution, the hot optically thin one, and a new cycle begins. 

In this scenario, we assume that the disk keeps the same magnetization level during the entire outburst. This means that, although the radial dependence of $B_z$ is fixed, the absolute value of $B_z$ needs to adjust to the evolving accretion rate. Magnetic rearrangements are expected to happen on a timescale that is at least as fast as the accretion timescale \citep{jacquemin2021} so this is not an issue. But in our scenario, one needs to keep a strongly magnetized disk even at low $h_\mathrm{th}/r$, which has historically been known to be a potential issue \citep{lubow1994a}. However, every simulations of magnetized thin disks up to now \citep{avara2016,liska2018b,zhu2018,mishra2020,liska2022,scepi2024} have shown that thin disks are able to retain a strong magnetic field even when going to $h_\mathrm{th}/r$ as low as $0.03$. Still, the regime of $h/r\approx 10^{-3}$ that we reach in the soft state remains inaccessible to numerical simulations and so our hypothesis that thin disks can be highly magnetized throughout an entire outburst remains to be checked with future numerical simulations. 

We have focused here on the simplest features of the hysteresis cycle of XRBs. However, many XRBs show more complex behaviors such as failed transitions to the soft state or different transition luminosities from the hard to the soft state in a given object. These complex behaviors are likely to involve the physics of the entire disk, e.g., how the outer disk feeds the inner disk with magnetic field and impacts its magnetization. Explaining those features is beyond the scope of this paper where we focus only on the physics of the inner disk. 

\section{Discussion}

\subsection{Comparison to the literature}\label{sec:comparison}
Very few simulations in the geometrically thin, magnetized regime exist for comparison with this work. Nonetheless, our present work is, to some extent, in agreement with the recent radiative GRMHD simulation of \cite{liska2022}, which shows that a thin disk at $L=0.3\:L_\mathrm{Edd}$ naturally goes to a temperature of $10^9$ K when it is strongly magnetized. This hot solution is allowed because, as in the simulation of \citetalias{scepi2024}, the density is much lower than in standard theory. However, we find that our simulation should be one-temperature at high accretion rate, in contrast to \cite{liska2022}. Moreover, we argue from our present analysis that there should be another solution that is optically thick at the same luminosity. This solution, however, would be hard to obtain from a radiative GRMHD simulation since it requires one to start from a very cold disk. Numerical simulations are now restricted, because of resolution, to aspect ratios of $h_\mathrm{th}/r\approx0.03$ while a temperature of $10^7$ K corresponds to $h_\mathrm{th}/r\approx10^{-3}$, an order of magnitude lower than what is currently feasible.

We have already pointed out the similarities between MAD numerical simulations and the JED semi-analytical solutions of \cite{petrucci2008,petrucci2010,marcel2018a,marcel2018b,marcel2019}. On \autoref{fig:hysteresis}, we plot our thermal solutions for our MAD simulation over a two-temperature JED solution with similar properties, i.e., a sonic Mach number of $m_s=0.5$ and a $\beta_{z,\:\mathrm{mid}}=20$. We see that the JED solution has two thermal solutions in the range $6\times10^{-5}\lesssim \dot{m} \lesssim 2\times 10^{-2}$, roughly an order of magnitude lower than in our case. We believe that this difference can be attributed to two effects: 

1) The effective scale height of the disk in our MAD solution is set to $3 h_\mathrm{th}$ (due to the additional support of magnetic pressure) while in the JED it is set to $h_\mathrm{th}$. Given that the surface densities between the JED and the MAD hot solutions are roughly similar (see \autoref{fig:hysteresis}), this means that our MAD solution is three times less dense than the JED for a given accretion rate.\footnote{Note that the inclusion of a turbulent magnetic pressure in JED solutions has been recently done by \cite{zimniak2024} (submitted).}

2) There is an additional source of energy coming from the innermost part of the disk in the MAD solution compared to the JED solution. To see this additional energy input, we refer the reader to \autoref{appendix:Appendix} where we analyze in detail the energy balance of our MAD simulation. Briefly, we find that, in our MAD solution, there is roughly twice the local binding energy that is available at large radii thanks to magnetic stresses that redistribute energy outward from the innermost radii (including the plunging region). In contrast, in the JED framework, it is assumed that only the local binding energy is available at each radius \citep{marcel2018a}. This comes from the neglect of GR effects and the related assumption of a Newtonian Keplerian rotation law, forbidding the existence of an innermost plunging region leading to radially oriented magnetic field lines, able to transfer angular momentum efficiently in the radial direction.

\subsection{A dark jet?}

A prominent feature of the hard-to-soft state transition is the disappearance of the radio emission, which is associated with a compact jet \citep{mirabel1994,fender1999,tetarenko2017}. Upper limits on the radio flux in 4U 1957+11 show that the radio emission from the jet is reduced by at least 2.5 orders of magnitudes when entering the soft state \citep{russell2011}. This is why many scenarios advocate for a change in the magnetization as the driver of the hard-to-soft state transition \citep{ferreira2006,igumenshchev2009,begelman2014}. Here we argue that, if the XRB hysteresis is driven by a thermal hysteresis, a high magnetization must be present all along the XRB outburst to explain the state transition luminosities. Hence, we would expect a powerful Blandford-Znakek jet \citep{blandford1977} to be present all along the outburst since it is an unavoidable outcome of every MAD simulation. However, we argue that it is not clear that this jet would be visible all along the outburst for two reasons. First, \citetalias{scepi2024} find that the efficiency of the BZ jet goes as $h_\mathrm{th}/r$ so that the radio emission could be expected to decrease by a factor of as much as $100$ when going from the hard state to the soft state. Note that \cite{avara2016} suggests an even steeper dependence of the jet efficiency that goes as $(h_\mathrm{th}/r)^2$. This is marginally enough to explain the quenching of the jet. The second point is that the power of the BZ jet is related in a non-trivial way to the radio emission that is observed. Particles need to be accelerated somehow and there is no consensus regarding the origin of the particle acceleration in jets. The internal shock model of \cite{malzac2013,malzac2014} suggests that particles are accelerated inside shocks due to to the differential jet velocity. One prediction of this internal shock model is the presence of a ``dark,'' weakly dissipative jet in the soft state \citep{drappeau2017,peault2019}, which would fit well in our scenario.\footnote{Note that in the original idea of the internal shock model, it is the fluctuations in velocity in a Blandford-Payne type jet that are responsible for the shocks. These fluctuations are themselves driven by the fluctuations in the accretion rate in the disk. Here, we implicitly assumed that the relation between the jet and the accretion flow is the same whether the jet is driven by a Blandford-Znajek or Blandford-Payne mechanism.}  Other models invoking particle acceleration in the sheath of the jet at the interface with the wind/accretion flow \citep{sironi2021} could produce a weakly dissipative jet in the soft state. Indeed, given that the disk is much thinner in the soft state, it would be natural to expect the interaction between the jet sheath and the wind/accretion flow to be much weaker in the soft state than in the hard state. Recent improvements in radio sensitivity with SKA might allow revisiting the lower radio detection limits and test again the presence of a weak jet in the soft state of XRBs.

\subsection{Application to AGN}

\begin{figure}
\includegraphics[width=85mm]{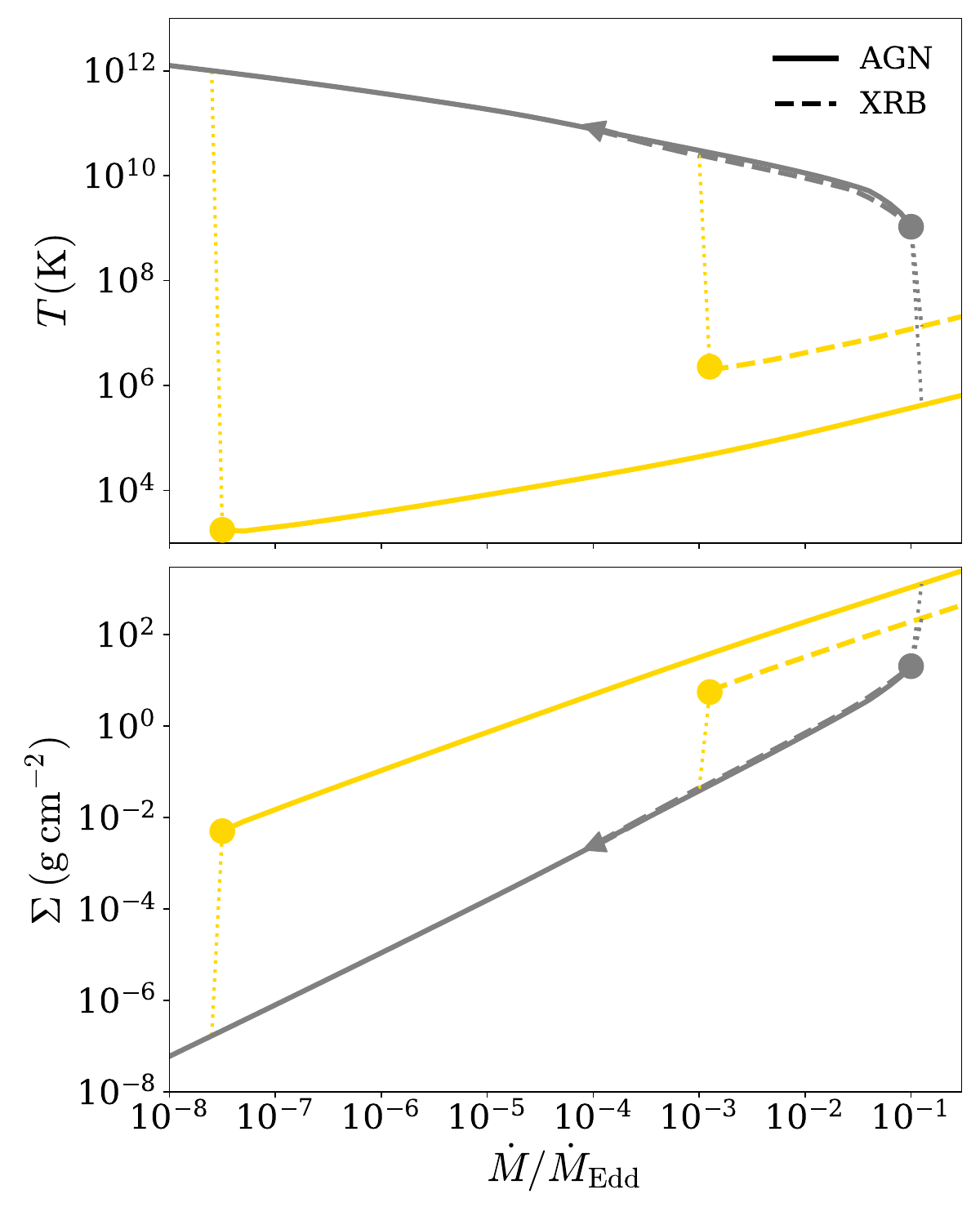}
\caption{Comparison of the thermal solutions for the MAD numerical simulation at $r=8\:r_g$ in an AGN with $M_{\rm BH}=10^8 M_\odot$ (solid lines) and an XRB with $M_{\rm BH}=10 M_\odot$ (dashed lines). The grey and gold lines respectively show the hot and cold branches. The hysteresis cycle for AGN extends over a much wider range of $\dot{m}$ so that low-luminosity AGN with $\dot{m}$ as low as $10^{-7}$ could be expected to be found in a disk-like spectral state.}
\label{fig:AGN_curves}
\end{figure}

It is often argued that AGN are scaled-up versions of XRBs so that the population of AGN should imprint the properties of the XRB hysteresis \citep{kording2008,ruan2019,fernandez2021,hagen2024}. To test this hypothesis, we apply our post-processing procedure to find the thermal equilibrium curves of an AGN around a BH of $M_{\rm BH}=10^8 M_\odot$. \autoref{fig:AGN_curves} shows the comparison of the thermal hysteresis drawn by an AGN (solid line) compared to an XRB (dashed line). The AGN case has a thermal hysteresis over a much wider range of $\dot{m}$ going from $ 3\times 10^{-8} \lesssim \dot{m} \lesssim 0.1$, compared to $ 10^{-3} \lesssim \dot{m} \lesssim 0.1$ for the XRB case. While the hot, optically thin branch has its highest $\dot{m}$ of equilibrium at $\approx 0.1$ in both the AGN and XRB cases, the cold, optically thick branch has its lowest $\dot{m}$ of equilibrium at $\approx 3\times 10^{-8}$ in the AGN case, i.e. much lower than the $\dot{m}\approx 10^{-3}$ for the XRB case. This difference arises from the fact that on the cold, optically thick branch the temperature of an AGN is much lower than that of an XRB. Since the Rosseland opacity depends so drastically on temperature, this means that for an AGN the density has to decrease by a very large amount compared to an XRB for the disk to become optically thin on the cold branch. 

In our thermal hysteresis scenario, we would therefore not expect AGN and XRB to have the same hysteresis cycle. The properties of an individual object in the hard state are expected to be similar in an AGN and an XRB, for example the anti-correlation between the X-ray hardness and the luminosity above $10^{-2}\:L_{\rm Edd}$. However, in contrast to XRBs we would expect to find that low-luminosity AGN with $\dot{m}\gtrsim 10^{-7}$ should be divided quite evenly between objects with a hard state and a soft state spectrum. The exact ratio of hard to soft state objects at a given luminosity will depend on the ratio of the time spent during the rise to outburst compared to the time spent during the decay to quiescence. This ratio will in turn depend on the mechanism driving the AGN eruptions. The ionization instability model, which has been successfully applied to XRBs, fails to explain the AGN population properties and does not produce outbursts similar to the XRB case \citep{hameury2009}. A mechanism involving gravitational instability in the outer disk might also be at work and so it is unclear what the secular outburst of an AGN should look like. Moreover, the few changing-look AGN that are observed to change spectral states evolve on time scales that are much shorter than a scale-up of the XRB time scales. This makes it hard for us to give a precise number on the expected fraction of low-luminosity AGN in a soft disk-like state. Nonetheless, this fraction should be higher than in XRBs and we do not expect to retrieve an observed lack of soft-state AGN below $10^{-2}\:L_{\rm Edd}$, as usually expected in analogy to XRBs. While recent observational surveys by \cite{fernandez2021} and \cite{hagen2024} have suggested that all AGN do transit to a hard state below $10^{-2}$ Eddington, we note that a recent X-ray sky-survey by eROSITA has found a lack of soft X-ray emitting AGN compared to what was expected from models based on the XRB analogy \citep{arcodia2024}. Clearly, this crucial point deserves more study on the theoretical and observational fronts.

\section{Conclusions}

We have studied the thermal properties of wind-driven accretion disks that are magnetically dominated ($\beta_{z,\:\mathrm{mid}}<100$), recognized as MADs in GRMHD simulations or as JEDs in analytical studies. By including radiative cooling in post-processing in a simulation from \citetalias{scepi2024}, we solve for all the equilibrium temperatures of the disk at different accretion rates along an XRB outburst.

We find that a MAD can maintain a hot inner disk (with temperatures $\gtrsim 10^{8-9}\:\mathrm{K}$) up to luminosities as large as $0.2\:L_{\rm Edd}$ (for an accretion efficiency of $\approx0.2$) potentially explaining the presence of a high-luminosity hard state in XRBs. This hot inner disk is inhomogeneous with cold patches of dense gas in the midplane. The hot inner disk cools from outside-in as the luminosity increases, yielding a ``truncation radius'' (i.e. a transition radius between a hot optically thin inner disk and a cold optically thick outer disk) that moves inward with increasing luminosity. 

By assuming simple scalings for the disk scale height and accretion speed with temperature, we find that, magnetically arrested disks have two thermal equilibrium solutions in the regime of $0.001\lesssim \dot{m} \lesssim 0.1$ at $r=8\:r_g$ and $0.04\lesssim \dot{m} \lesssim 0.5$ at $r=3\:r_g$. This provides the means for a thermal hysteresis in XRBs without invoking a strong ADAF-principle as is often done. Starting from the quiescent state, the disk stays in the hard state until there is no hot solution anymore. This happens around $0.2\:L_{\rm Edd}$ at $r=8\:r_g$ where the disk goes to the only available solution, the cold branch. Going down to quiescence, the disk then stays on the cold solution until it ceases to exist. At $8\:r_g$, this happens at luminosities of $\approx 0.002\:L_\mathrm{Edd}$. Within our assumptions, our transition luminosities between the cold and hot solutions match reasonably well the transitions observed in the hysteresis of XRBs. In fact, we find that our transition luminosities are $6$ times higher than in the case of the  Newtonian JED semi-analytical solution of \cite{marcel2018a} because of the effect of magnetic pressure and additional energy transport, mediated by magnetic stresses, from the innermost part of the disk below the ISCO. 

Finally, we emphasize that in our case, the spectral hysteresis is purely thermal and the disk is always strongly magnetized. This means that we expect a jet, although possibly weakly dissipative, to be present even in the soft state. We also naively extrapolate our analysis to the case of AGN and find that the hysteresis would happen over a much larger range of accretion rate, i.e. for $10^{-7}\lesssim \dot{m} \lesssim 2\times10^{-1}$. This means that, under the hyopthesis that the structure of the inner disk of an AGN is the same as that of an XRB, we should expect a significantly larger fraction of low-luminosity AGN to be in a soft state compared to the XRB case.

\begin{acknowledgements}
We thank the referee for their constructive report that improved siginificantly the clarity of our paper. NS thanks Eliot Quataert for fruitful discussions on an early version of the paper and for suggesting that we extend our scenario to AGN. NS acknowledges partial financial support from the European Research Council (ERC) under the European Union Horizon 2020 research and innovation programme (Grant agreement No. 815559 (MHDiscs)) and the UK's Science and Technology Facilities Council [ST/M001326/1]. We also acknowledge financial support from  NASA Astrophysics Theory Program grants NNX16AI40G, NNX17AK55G, 80NSSC20K0527, 80NSSC22K0826 and an Alfred P. Sloan Research Fellowship (JD). GM acknowledges financial support from the Polish National Science Center grant 2023/48/Q/ST9/00138.
\end{acknowledgements}

%
%

 \bibliographystyle{aa} 
 \bibliography{biblio}

\begin{appendix}

\section{Energy balance in a MAD simulation}\label{appendix:Appendix}
We start by recalling that the continuity equation and the conservation of energy in GRMHD read as
\begin{gather}
    \nabla_\mu (\rho u^{\mu})  = 0,\label{eq:GRMHD_continuity}\\
    \nabla_\mu T^{\mu}_{t} + \mathcal{F}_\mu = 0,\label{eq:GRMHD_energy}
\end{gather}
where the stress-energy tensor is
\begin{equation}
    T^{\mu}_{\nu} = (\rho + \gamma_\mathrm{ad}u_g +b^\lambda b_\lambda)u^\mu u_\nu -b^\mu b_\nu
\end{equation}
and $\mathcal{F}_\mu$ represents an optically thin cooling source term that is defined as in \cite{noble2009}. Additionally, $p_\mathrm{gas}$ is the gas pressure, $u_g=p_\mathrm{gas}/(\gamma_\mathrm{ad}-1)$ is the internal energy of the gas in the comoving frame, $\gamma_\mathrm{ad}=5/3$ is the adiabatic index and $b^\mu$ is the fluid frame 4-magnetic field. 

\begin{figure}
\includegraphics[width=80mm]{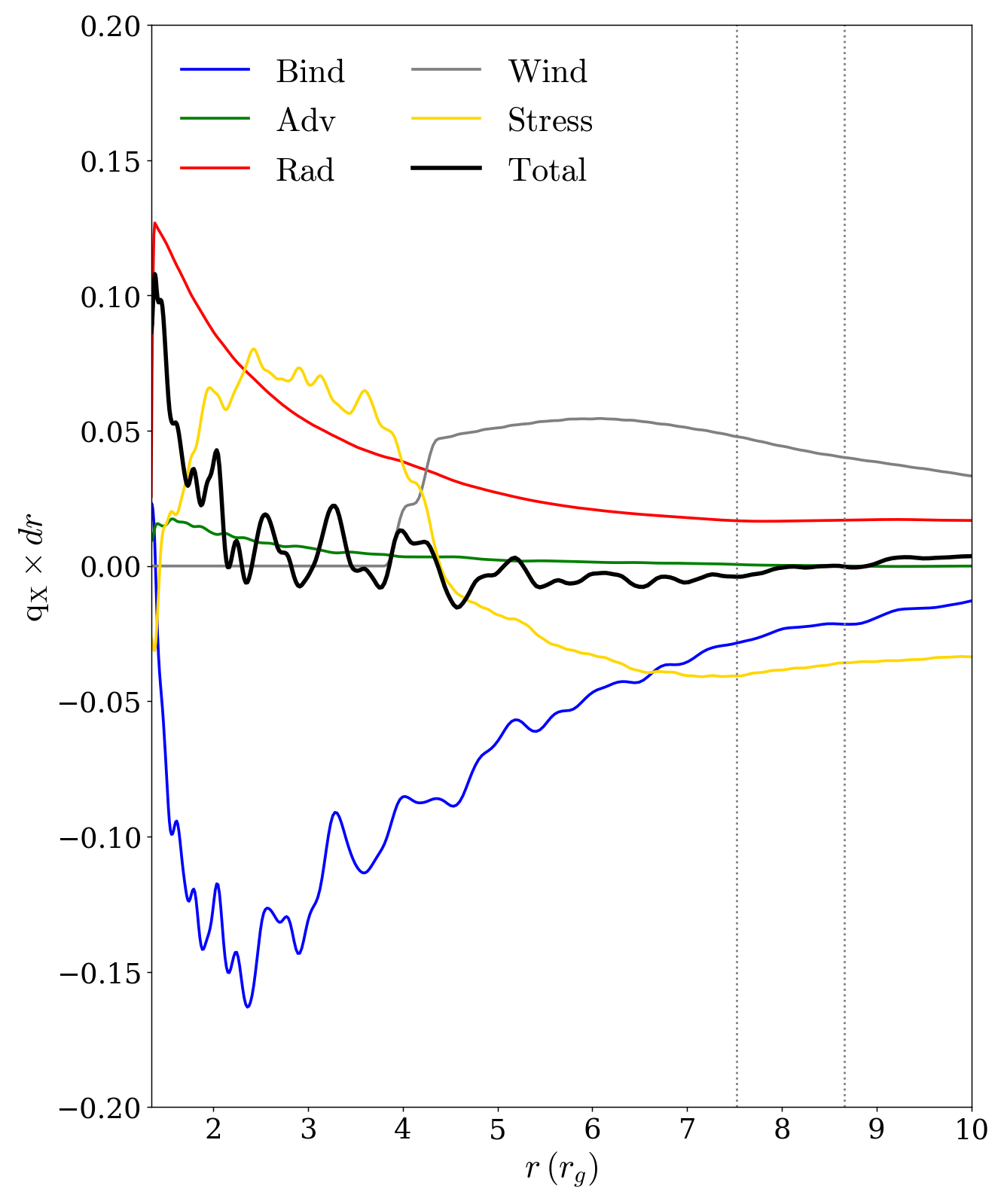}
\caption{Balance of energy loss/gain in the disk as a function of radius. The blue, green, red, grey and gold lines show respectively $q_\mathrm{{Bind}}$ (the binding energy gain), $q_\mathrm{{Adv}}$ (the disk internal energy advected energy), $q_\mathrm{{Rad}}$ (the radiative energy loss), $\mathrm{L_{Wind}}$ (energy loss due to the wind) and $q_\mathrm{{Stress}}$ (energy loss/gain due to stresses in the disk). $q_\mathrm{{Stress}}$ changes sign around $4.5\:r_g$ meaning that energy is redistributed from the inner region to the outer regions of the disk by magnetic stresses.}
\label{fig:energy_fluxes}
\end{figure}

We then define a surface delimiting the boundary between the disk and the wind. For simplicity, we define this boundary at a fixed $\theta$ to have only fluxes along $\theta$ in the wind contribution and fluxes along $r$ in the disk contribution. We define this boundary at $\theta_\mathrm{\pm}=\pi/2 \pm \pi/8$. We have checked that our choice of the disk surface is close to the surface where the angular momentum flux, $T^r_\phi$, changes sign \citep{ferreira1995,manikantan2024}. We then integrate the sum of \autoref{eq:GRMHD_continuity} and \autoref{eq:GRMHD_energy}\footnote{We add the continuity equation to remove the rest-mass energy contribution (see \citealt{sadowski2016}).} over the volume $[t_\mathrm{beg},t_\mathrm{end}]\times[\theta_+,\theta_-]\times[0,2\pi]$ in the $t,\theta,\phi$\footnote{$t_\mathrm{beg}$ and $t_\mathrm{end}$ are defined in \citetalias{scepi2024}} directions  and we assume steady-state conditions to define $q_\mathrm{tot}$, the total energy loss in the disk, 
\begin{equation}\label{eq:GRMHD_eq}
    q_\mathrm{tot}\equiv \int^{\theta^+}_{\theta^-} \partial_r(\sqrt{-g} \langle T^{r}_{t} \rangle )d\theta + \big[ \sqrt{-g} \langle T^{\theta}_{t} \rangle  \big]^{\theta^+}_{\theta^-} + \int^{\theta^+}_{\theta^-}  \sqrt{-g}\langle \mathcal{F}_t \rangle d\theta = 0,
\end{equation}
where $\langle \rangle$ means an azimuthal and time average. We see that by construction, $q_\mathrm{tot}$ should be equal to 0. This can be seen to be approximately true \footnote{Note that radial derivatives introduce a lot of noise in the procedure.} on \autoref{fig:energy_fluxes} except near the horizon where the heavy use of ceilings on the magnetic field breaks energy conservation. 

To gain physical insight, we decompose $q_\mathrm{tot}$ into five type of energy losses that reduce to well-known quantities in a Newtonian framework. With this decomposition, we have
\begin{equation}\label{eq:tot_energy}
    q_\mathrm{tot} = q_\mathrm{Rad}+q_\mathrm{Bind}+q_\mathrm{Adv}+q_\mathrm{Stress}+q_\mathrm{Wind}=0,
\end{equation}
with
\begin{gather}
    q_\mathrm{Rad} = \langle \int_{\theta} \sqrt{-g}\mathcal{F}_t d\theta\rangle,\\
    q_\mathrm{Bind} =  \int_{\theta} \partial_r(\sqrt{-g}\langle \rho u^r \rangle \langle u_t \rangle) d\theta,\\
    q_\mathrm{{Adv}} =  \int_{\theta} \partial_r(\sqrt{-g}\langle \gamma_\mathrm{ad}u_g u^r \rangle \langle u_t \rangle) d\theta,\\
    q_\mathrm{Stress} =  \int_{\theta} \partial_r(\sqrt{-g}\langle \delta([\rho+\gamma_\mathrm{ad}u_g] u^r) \delta u_t + b^2 u^ru_t -b^rb_t \rangle) d\theta,\\
    q_\mathrm{Wind} =  \big[ \sqrt{-g}\langle (\rho+\gamma_\mathrm{ad}u_g+b^2) u^\theta u_t -b^\theta b_t  \rangle  \big]^{\theta_\mathrm{+}}_{\theta_\mathrm{-}}.
\end{gather}
Here $\mathrm{q_{Rad}}$ is the radiative energy loss, $\mathrm{q_{Bind}}$ is the binding energy loss, $\mathrm{q_{Adv}}$ is the disk internal energy advection energy loss, $\mathrm{q_{Stress}}$ is the energy loss due to stresses through the disk and $\mathrm{q_{Wind}}$ is the energy loss due to the wind. Note that to define $q_\mathrm{{Bind}}$, $q_\mathrm{{Adv}}$ and $q_\mathrm{{Stress}}$ we divided the $\langle (\rho+\gamma_\mathrm{ad}u_g)u^ru_t\rangle$ into a laminar component $\langle (\rho+\gamma_\mathrm{ad}u_g)u^r\rangle \langle u_t\rangle$ (going into $q_\mathrm{{Bind}}$ and $q_\mathrm{{Adv}}$) and a turbulent component $\langle \delta([\rho+\gamma_\mathrm{ad}u_g]u^r)\delta u_t\rangle\equiv \langle  (\rho+\gamma_\mathrm{ad}u_g)u^ru_t \rangle - \langle (\rho+\gamma_\mathrm{ad}u_g)u^r\rangle \langle u_t\rangle$ (going into $q_\mathrm{{Stress}}$).

We plot on \autoref{fig:energy_fluxes} these five contributions. Note that a positive (negative) energy loss means that the disk effectively loses (gains) energy. For example, $q_\mathrm{{Rad}}$, $q_\mathrm{{Adv}}$ and $q_\mathrm{{Wind}}$ are positive, meaning that radiation, advection and the wind remove energy from the disk. In fact, advection is almost negligible here but the wind takes a large part of the energy compared to radiation. On the other hand, $q_\mathrm{{Bind}}$ is negative meaning that binding energy is deposited in the disk. Now, $q_\mathrm{{Stress}}$ changes sign around $4.5\:r_g$ going from positive to negative. This means that energy is redistributed from the inner region to the outer regions of the disk by magnetic stresses. This is similar to what happens in a standard model using a zero-torque boundary condition at the horizon of the black hole (but not at the ISCO as usually assumed). We find that this redistribution of energy can induce a local energy gain that can be as much as 3.5 times the local binding energy gain at $r=10\:r_g$. 

To be more quantitative, we look at the difference of the luminosities between $7.5\:r_g$ and $8.5\:r_g$ to see how much energy is deposited/taken away at $r=8\:r_g$ (the radius at which \autoref{fig:hysteresis} is made). We write $\Delta E_X = \int_{7.5\:r_g}^{8.5\:r_g}q_X dr$. As such, we can rewrite \autoref{eq:tot_energy} in a form close to Eq. 10 of \cite{petrucci2010},
\begin{equation}
-\Delta \mathrm{E_{Bind}} - \Delta \mathrm{E_{Stress}} = \Delta \mathrm{E_{Rad}} + \Delta \mathrm{E_{Adv}} + \Delta \mathrm{E_{Wind}},
\end{equation}
where $-\Delta \mathrm{E_{Stress}}$ represents an additional source of energy, on top of the local binding energy, compared to the Eq. 10 of \cite{petrucci2010}.

We find that $\Delta \mathrm{E_{Rad}}\approx 0.33$, $\Delta \mathrm{E_{Bind}}\approx -0.48$, $\Delta \mathrm{E_{Adv}}\approx 0.005$, $\Delta \mathrm{E_{Stress}}\approx -0.77$ and  $\Delta \mathrm{E_{Wind}}\approx 0.90$. Again, this means that roughly 2.5 times the equivalent of the binding energy is locally available at each radii. Out of this local energy roughly three quarters is taken out by the wind while only one quarter is locally radiated away. 

It is intriguing that $q_\mathrm{{Stress}}$ is not zero at the ISCO as this means that magnetic stresses could potentially tap into the rotational energy of the black hole inside the ergosphere. To check this idea, we have run a preliminary thin ($h_\mathrm{th}/r=0.03$ as in \citetalias{scepi2024}) MAD simulation with zero spin. We find that $q_\mathrm{{Stress}}$ is similar in the cases with zero and high spin, disfavoring the rotational energy of the black hole as a source of energy here. Clearly, this point deserves more work and will be the subject of a separate paper (Scepi et al. in preparation), in which we will present the details of a simulation with zero spin.

\section{Decomposition of the cooling mechanisms}\label{appendix:AppendixB}

\begin{figure}
\includegraphics[trim={0 50mm 0 110mm},clip=true,width=90mm]{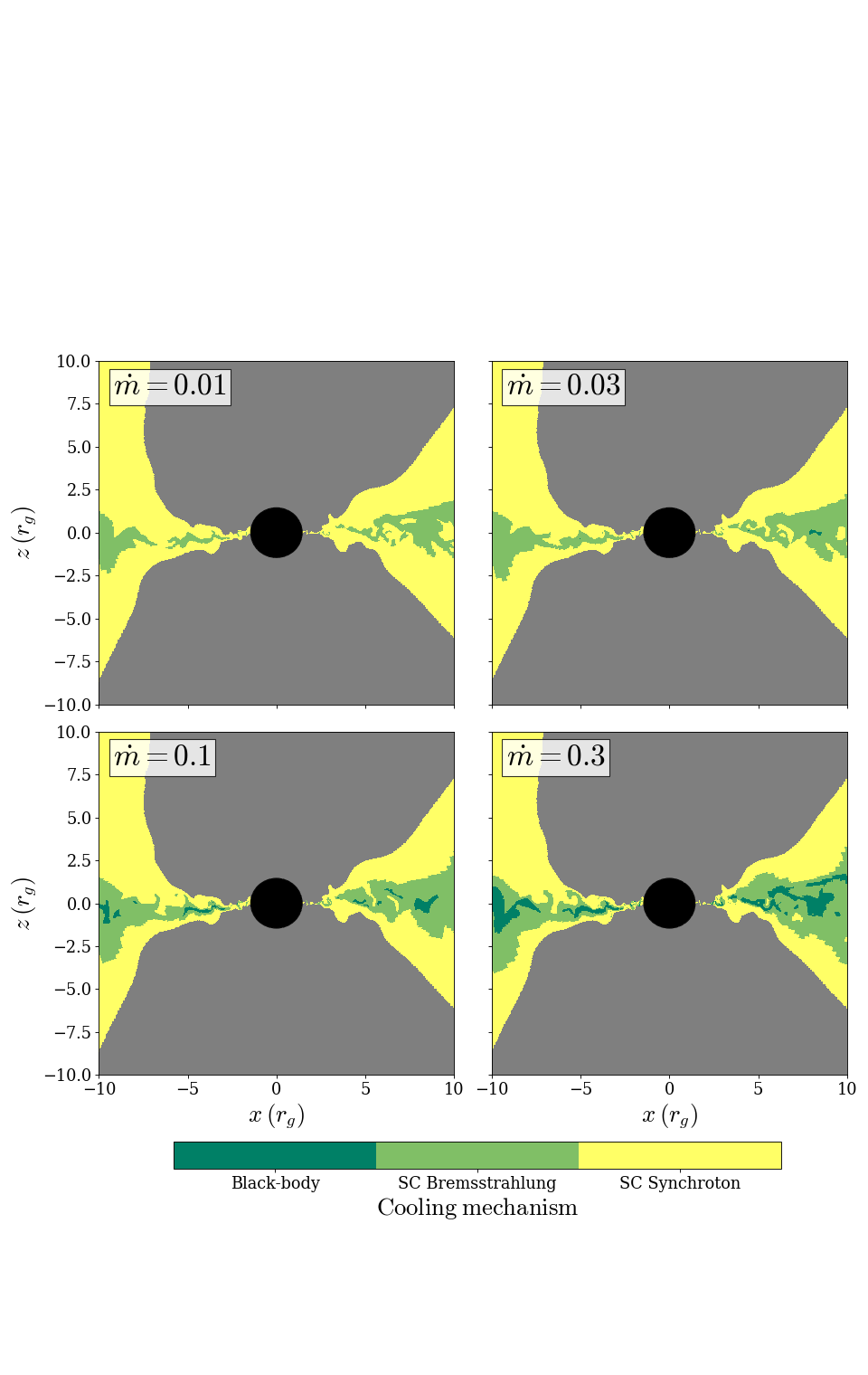}
\caption{Poloidal cut showing the dominant cooling mechanism in each cell for four different $\dot{m}$. The grey region shows the jet region (defined as the region where $\sigma>1$) that is excluded from our analysis. The snapshot used is the same as in \autoref{fig:midplane_T_hard} to allow comparison of the two Figures.}
\label{fig:cooling_poloidal}
\end{figure}

\begin{figure}
\includegraphics[trim={0 50mm 0 110mm},clip=true,width=90mm]{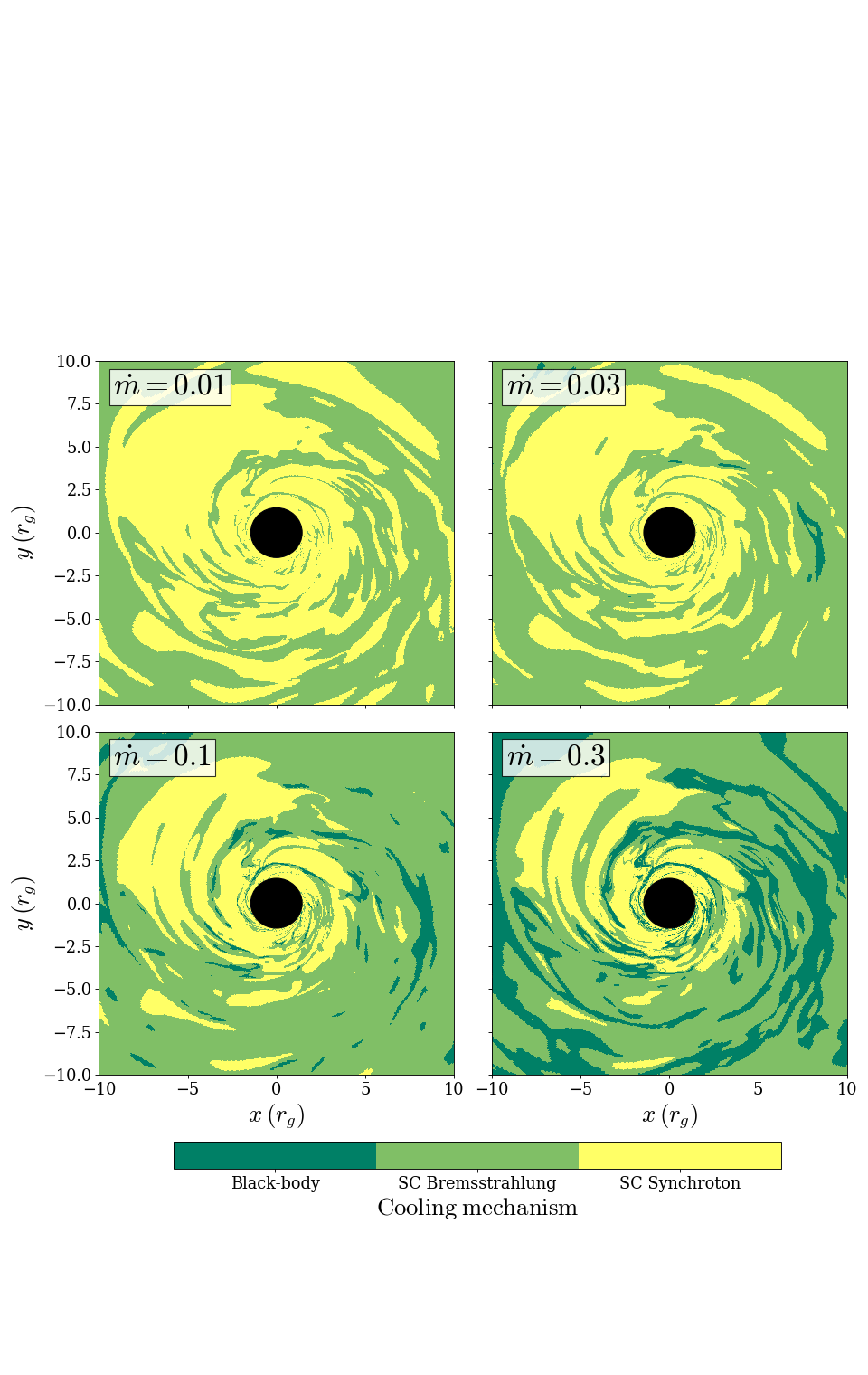}
\caption{Midplane cut showing the dominant cooling mechanism in each cell for four different $\dot{m}$. The snapshot used is the same as in \autoref{fig:poloidal_T_hard} to allow comparison of the two Figures.}
\label{fig:cooling_midplane}
\end{figure}

\autoref{fig:cooling_poloidal} and \autoref{fig:cooling_midplane} respectively show the dominant cooling mechanisms in each cell of the grid in a poloidal cut and a midplane cut of the disk. The snapshots are the same as used in \autoref{fig:poloidal_T_hard} and \autoref{fig:midplane_T_hard} to allow comparisons with the temperature maps. At low accretion rates, we see that the disk is cooling entirely through optically thin processes (synchrotron self-Compton (SSC) and bremsstrahlung self-Compton (BSC)). The inner parts of the disk are cooling mostly through SSC whereas the outer parts are cooling mostly through BSC. This is because the magnetic field goes as $r^{-1.5}$ whereas the density goes as $r^{-1}$. Because of their low density, the upper layers of the disk are also dominated by SSC emission. As the accretion rate increases and the density increases, we see that BSC starts to gradually take over SSC in the inner disk. We also see that the densest part of the disk start to cool through optically thick black-body emission to form clumps of cold materials (as can be seen in \autoref{fig:poloidal_T_hard} and \autoref{fig:midplane_T_hard}.)

\end{appendix}

\end{document}